\documentclass[useAMS,usenatbib]{mnras}

\usepackage[T1]{fontenc}
\usepackage{ae,aecompl}

\usepackage{amsfonts}
\usepackage{bbold}
\usepackage{color}
\usepackage{transparent}
\usepackage{transparent}
\usepackage{rotating}
\usepackage{caption}
\usepackage{subcaption}

\usepackage{graphicx}	
\usepackage{amsmath}	
\usepackage{amssymb}	
\usepackage{tablefootnote}
\usepackage[flushleft]{threeparttable}

\usepackage{xspace}
\newcommand{\sgx}{Sg\textsc{xb}\xspace}
\newcommand*{\eg}{e.g.\@\xspace}
\newcommand*{\ie}{i.e.\@\xspace}
\newcommand*{\aka}{a.k.a.\@\xspace}
\newcommand*\diff{\mathop{}\!\mathrm{d}}
\newcommand*{\rhs}{r.h.s.\@\xspace}
\newcommand*{\lhs}{l.h.s.\@\xspace}
\newcommand*{\rlof}{\textsc{rlof}\@\xspace}

\newcommand*{\hmxb}{\textsc{hmxb}\@\xspace}

\newcommand*{\sfxt}{\textsc{sfxt}\@\xspace}

\newcommand*{\ns}{\textsc{ns}\@\xspace}

\newcommand*{\msun}{$M_{\odot}$\@\xspace}
\newcommand*{\rsun}{$R_{\odot}$\@\xspace}

\newcommand*{\bhl}{\textsc{bhl}\@\xspace}

\newcommand*{\cak}{\textsc{cak}\@\xspace}

\newcommand*{\kms}{km$\cdot$s$^{-1}$\@\xspace}

\title[Wind in \sgx. I - Orbital structure]{A numerical investigation of wind accretion\\ in persistent Supergiant X-ray Binaries\\I - Structure of the flow at the orbital scale}

\author[I. El Mellah and F. Casse]{I. El Mellah$^{1}$\thanks{\href{mailto:ileyk@apc.univ-paris7.fr}{ileyk@apc.univ-paris7.fr}} \& F. Casse$^{1}$\\
$^{1}$APC AstroParticule \& Cosmology laboratory - Universit\'e Paris 7 Diderot Sorbonne Paris Cit\'e \\ 10 rue Alice Domon et L\'eonie Duquet, Paris, 75013, France}

\date{Accepted XXX. Received YYY; in original form ZZZ}

\pubyear{2016}

\begin{document}
\label{firstpage}
\pagerange{\pageref{firstpage}--\pageref{lastpage}}
\maketitle

\begin{abstract}
Classical Supergiant X-ray Binaries host a neutron star orbiting a supergiant OB star and display persistent X-ray luminosities of $10^{35}$ to $10^{37}$\,erg$\cdot$\,s$^{-1}$. The stellar wind from the massive companion is believed to be the main source of matter accreted by the compact object. With this first paper, we introduce a ballistic model to characterize the structure of the wind at the orbital scale as it accelerates, from the stellar surface to the vicinity of the accretor. Thanks to the parametrization we retained and the numerical pipeline we designed, we can investigate the supersonic flow and the subsequent observables as a function of a reduced set of characteristic numbers and scales. We show that the shape of the permanent flow is entirely determined by the mass ratio, the filling factor, the Eddington factor and the $\alpha$-force multiplier which drives the stellar wind acceleration. Provided scales such as the orbital period are known, we can trace back the observables to evaluate the mass accretion rates, the accretion mechanism, the shearing of the inflow and the stellar parameters. We confront our model to three persistent Supergiant X-ray Binaries (Vela X-1, IGR J18027-2016, XTE J1855-026) and discuss the likelihood of wind-formed accretion discs around the accretors in each case, further investigated in a following paper.
\end{abstract}




\section{Introduction}

For the last fifty years, X-ray observations have revealed a plethora of binary systems hosting a compact object accreting the gas from its stellar companion. While accretion proceeds mostly by overflow of the stellar Roche lobe (the Roche Lobe OverFlow - \textsc{rlof}) in systems where the mass of the star is low (Low Mass X-ray Binaries, {\sc lmxb}), we expect the dense and fast winds of massive stars to be the main responsible for mass transfer in High Mass X-ray Binaries ({\sc hmxb}). The structure of the wind in {\sc hmxb} has been found to be of two kinds \citep{Chaty2011a} : either it forms a circumstellar decretion disc around fast rotators Be stars (Be{\sc xb}) or it obeys the more isotropic sketch of the radiatively-driven wind for early-type supergiant stars \citep{Shakura:2014wk}. The latter sources of matter are believed to be associated to the class of the persistent classical Supergiant X-ray binaries (henceforth \sgx) when surrounded by a neutron star on a low eccentricity orbit thus making the compact object permanently embedded in the wind. The number of confirmed \sgx has doubled within the last ten years \citep{Walter15}.

The theory of radiatively-driven winds (or \textsc{cak} winds) for isolated massive stars has been widely developed, refined and confronted to observations since the seminal papers of the 70's \citep{Lucy1970,Castor1975}. It describes how the resonant absorption of the high energy photons of hot stars by non fully ionized metals supplies the outer layers with net linear momentum. As they accelerate, they keep absorbing Doppler shifted photons previously untouched. In its most elementary form, steady-state solutions can be derived from 3 parameters, the force multipliers, which set the mass loss rate for a given mass and luminosity of the star. Radial velocity profiles have also been determined and have turned out to be well fitted by the so called $\beta$-laws. The terminal speeds involved are typically of the order of several escape velocities, which match the observationally derived values using a wide spectrum of methods summarized in the very comprehensive reviews of \cite{Kudritzki2000} and \cite{Puls2008}. This model has reached levels of reliability high enough, in particular in the case of supergiant stars \citep{Crowther2006}, so as to be confidently applied to the case of \sgx. 

In this paper, we focus on the self-consistency of the stellar, orbital, wind and accretion parameters in \sgx hosting a confirmed neutron star. Due to the high mass ratio in theses systems, we do not expect the star to fill its Roche lobe nor the launching of the wind to be altered by the presence of the neutron star ; yet, the non inertial forces at stake in those relatively short periods systems ($<20$ days) will modify the structure of the departed though still accelerating flow as it approaches the compact object, where the gravitational field of the latter will take over. Hence, an observable such as the spectral type of the supergiant plays a role in the determination of the force multipliers which determine the wind properties (\eg the mass loss rate), but also in shaping the Roche potential in which the modified velocity profile unfolds. In \sgx where the stellar companion has been identified, we usually rely on the spectral type to determine the stellar parameters, putting aside the specificities inherent to the binary secular evolution \citep{VandenHeuvel2009}. Still today, mass discrepancies subsist between the spectroscopic and evolutionary masses \citep{Searle2008,Puls2008a}. To those approaches we want to add measures derived from their consequences over the whole mass transfer phenomenon and considered in relation with each other through the launching mechanism, the orbital dynamics, the stellar model and the accretion process. Since \sgx are highly obscured systems \citep{Coleiro2013}, thorough observations of the star are challenging and sometimes impossible. Using accretion onto the compact companion could be used as a proxy to trace back information concerning the whole system, including the donor.

Radiatively-driven winds are notoriously unstable to small-scale radial perturbations \citep{MacGregor1979, Owocki1984, Owocki1985} and tend to form clumps of overdense matter. Their accretion has been suggested as a possible source of X-ray variability \citep{Ducci2009} but other phenomena are expected to play a role \citep{Illarionov1975,Foulkes2006,Shakura2013a}, in particular in recently discovered Super Fast X-ray Transients \citep[\textsc{sfxt,}][]{Negueruela2006} where the X-ray dynamic range spans up to 5 orders of magnitude. Apart from the off-states, the X-ray emission is more persistent in classical \sgx with averaged X-ray luminosities of the order of $10^{35}$ to a few $10^{36}$\,erg\,s$^{-1}$. In this first paper, we try to represent the averaged behaviour with a steady-state framework and do not consider the possible switches which could occur due to the inhibition of the wind acceleration once the X-ray emission from the gas being accreted is large enough to fully ionize the wind \citep{Ho1987,Blondin1990}. We assume that permanent regime streamlines of the supersonic flow can tell us something about the properties of the system, regardless of ionizing effects or shocks at the orbital scale. 

This steady-state assumption serves another purpose in the attempt to understand variability in \sgx (\eg the off-states) and possibly in \textsc{sfxt}. The more regular and somewhat smoother accretion process at stake in \textsc{lmxb}, \textsc{rlof}, does not prevent those systems from undergoing important luminosity variations \citep[see \eg the Q-shaped diagram represented in Figure 2 of ][]{Belloni2005}, It turns out that phenomena occurring within the accretion disc itself could account for them. In the case of \sgx, highly obscured systems, we have few if not no proof of the existence of a disc-like structure around the accretor, except for a few ones which harbor fastly rotating pulsars such as Cen X-3 and are believed to undergo \rlof, in spite of the supergiant nature of the donor star. Some systems present observational signatures associated with direct impact at the poles where the matter could have been led by its interaction with the magnetosphere \citep{Davidson1973}. However, whether the gas is taped directly in a shocked wind or in a wind-formed disc, possibly transient, remains unknown. Thus, neglecting time-variability at the orbital scale is a way to decouple it from its small scale counterpart, in the vicinity of the neutron star where most of the X-ray emission comes from. Local triggers of instabilities in the shocked wind around the accretor do not necessarily require large scale excitation - whose damping or amplification remains largely unknown. \cite{Manousakis2015c} showed for instance that continuous homogeneous winds could lead to unstable front shocks due to X-ray ionizing feedback on the flow. The very stability of essentially planar supersonic accretion flows has been a long-lasting question \citep{Blondin2009,Blondin:2012vf} we want to address in the context of \sgx, in a similar way it has been addressed in the case of symbiotic binaries \citep{Theuns1993,Theuns1996,Jahanara2005,Mohamed2007,deValBorro:2009gk,HuarteEspinosa:2012wq}. 

We set the modelling stage of our toy-model in section \ref{sec:first_sec} and highlight its fundamental parameters. After a description of our numerical pipeline in section \ref{sec:second_sec}, we discuss the structure of the flow at the orbital scale and its properties (mass accretion rate onto the compact object and shearing of the flow) in section \ref{sec:structure_wind}. Eventually, in section \ref{sec:dim_res}, we evaluate the self-consistency of our cross-model by confronting the parameters we derive to three extensively studied \sgx : Vela X-1, XTE J1855-026 and IGR J18027-2016 (\aka SAX 1802.7-2017). Finally, we summarize our results and discuss their main implications in section \ref{sec:conclu}.

\section{Radiatively-driven wind in a SgXB Roche potential}
\label{sec:first_sec}
\subsection{Radiatively-driven winds of hot stars}
\label{sec:windModel}
\subsubsection{The bedrock model}
\label{bedrock}
In addition to the radiative pressure due to Thomson scattering of light on the free electrons in stellar atmospheres, spectral lines of partly ionized metals can induce a resonant absorption of the high energy photons provided by hot stars \citep[see \eg ][]{Lamers1999}. The underlying radiations tend to produce a net acceleration of those ions in the same direction, as they isotropically re-emit the absorbed photons from below - at least in the single scattering limit. Through Coulomb coupling of those metal ions with the ambient charged particles, momentum is redistributed and the whole gas is lifted up from the outer layers of the star. As the radial velocity rises, Doppler shifting of the absorption lines enables the metal ions to keep tapping radiative energy. Under additional assumptions concerning the spectral and spatial narrowness of the absorption process (the Sobolev approximation) and if we reduce the emission region to a point, one can write the corresponding acceleration term as \citep{Castor1975} :
\begin{equation}
\label{eq:CAK}
g_{\textsc{cak}}=\frac{Q}{1-\alpha}\cdot\frac{\kappa _e L_*}{4\pi r^2 c} \cdot \left( \frac{1}{\rho c \kappa _e Q } \frac{\diff v}{\diff r} \right)^{\alpha}
\end{equation}
One recognizes a contribution analogous to the radiative acceleration due to the continuum opacity by electron scattering (the second factor on the right hand side), with $\kappa _e$ being the free electrons opacity, $L_*$ the stellar luminosity, $c$ the speed of light and $r$ the distance to the stellar center. The $\alpha$ and $Q$ factors are the two dimensionless force multipliers, the fundamental parameters which structure the form and the amplitude of the physical quantities of the flow. The former stands for the relative contribution of thin and thick absorption lines to the net acceleration while the latter has been introduced by \cite{Gayley1995} in replacement of the initial $K$-factor of \cite{Castor1975} which was more correlated with $\alpha$. $Q$ is the quality factor corresponding to the coupling of radiation with the resonant spectral lines, is expected to lie between $500$ and $2,000$ and relates to $K$ through $K=Q^{1-\alpha}\left(v_{th}/c\right)^{\alpha}/(1-\alpha)$, with $v_{th}$ the thermal speed at the stellar surface. The third force multiplier, $\delta$, will not be considered in this paper\footnote{See \citealt{Puls2008} for a discussion on the relevance of $\delta$.} since its value is generally close to zero for the winds we consider \citep[it is the case of "frozen-in" ionization state discussed by ][in section 4.1]{Kudritzki1989a}. Notice that its inclusion would have lowered the terminal speed of the wind \citep[see section 5.2.2. in ][]{Groenewegen1989}. The values of $\alpha$, $Q$ and $\delta$ are computed from the opacity profiles of atomic lines and given for different stellar temperatures, surface gravity and metallicities. $\rho$ refers to the mass density of the flow which goes as $r^{-2}v^{-1}$ in this spherically symmetric framework. The velocity gradient in the last factor is originates from the aforementioned Doppler shifting. This expression is obtained under the assumption that photons undergo single scattering, potentially leading to slight underestimations of the mass outflow \citep[see \eg ][for proper handling of multiple scattering through Monte-Carlo methods]{Muller2008}. It also assumes a firm Coulomb coupling of the line-driven ions with the ambient gas, which is easily verified for the Sg stars we will consider in this paper.

\begin{figure*}
\begin{subfigure}{.5\textwidth}
  \centering
 \includegraphics[width=\columnwidth]{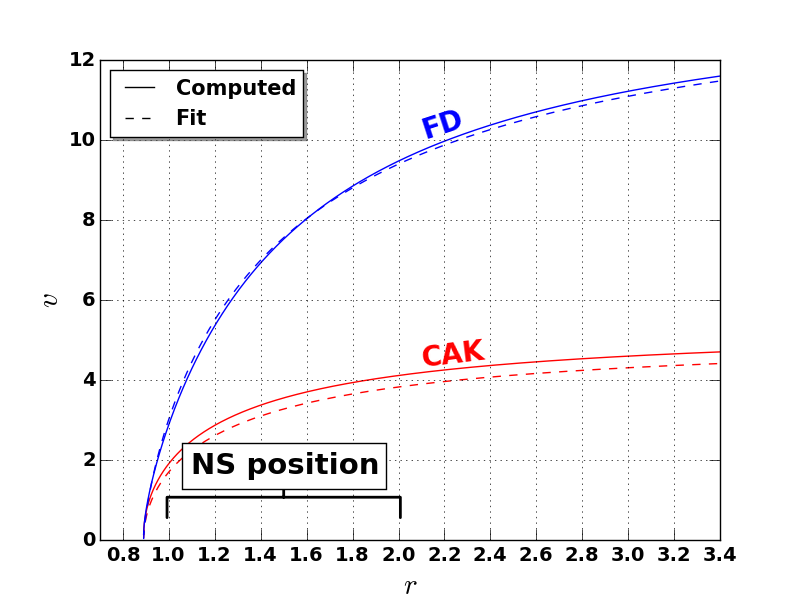}
  \label{fig:sfig1}
\end{subfigure}%
\begin{subfigure}{.5\textwidth}
  \centering
 \includegraphics[width=\columnwidth]{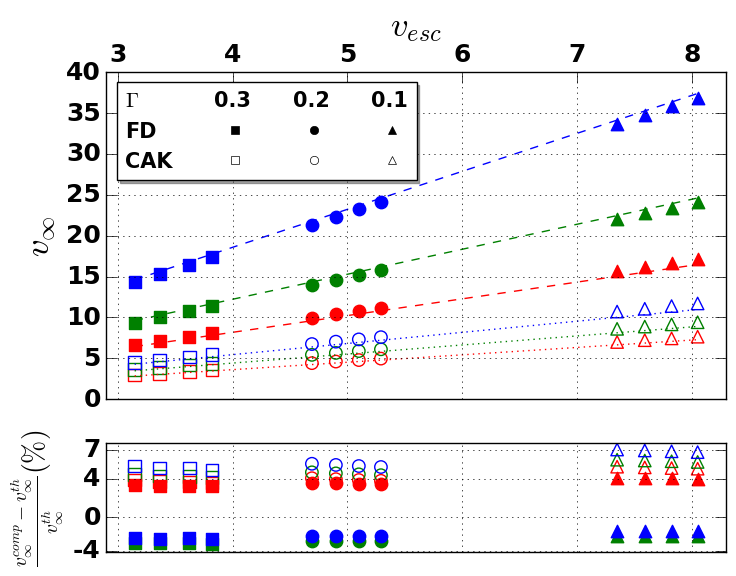}
  \label{fig:sfig2}
\end{subfigure}
\caption{Summarize of the preliminary check of the integrator in the case of the wind of an isolated star. (left panel) Velocity profile (in the arbitrary units system of section \ref{zemodel}) in the bedrock \cak model (red) and in the case where the finite disc effect has been included (blue). The red and blue dashed lines are the corresponding approximated profiles given respectively by equations \eqref{eq:vr_ana} and \eqref{eq:vr_FD}. (right panel) Terminal speeds (measured in our simulations beyond $r=30$) as a function of the effective escape velocity. Red, green and blue correspond respectively to $\alpha=0.45$, $0.55$ and $0.65$. $\Gamma$, $q$ and $f$ have been tuned so as to explore a wide range of escape velocities, still given in the units system described in the text. The relative differences between the computed and the expected values are represented below the main plot.}
\label{fig:1DCAK}
\end{figure*}

One can already notice that, in spite of the apparent cumbersome aspect of this most simple form of the \textsc{cak} wind, the only parameter which sets the shape of the flow is $\alpha$. Indeed, with tildes referring to adimensioned quantities, equation \eqref{eq:CAK} can be rewritten as :
\begin{equation}
\tilde{g}_{\textsc{cak}}= \frac{1}{\tilde{r}^{2(1-\alpha)}} \left( \tilde{v} \frac{\diff \tilde{v}}{\diff \tilde{r}} \right)^{\alpha}
\end{equation}
Thus, the steady state pressureless equation of motion of the flow is given by, once adimensioned with the relevant quantities (see below) :
\begin{equation}
\label{eq:CAKadim}
v\frac{dv}{dr}=-\frac{1}{r^2}+\frac{\aleph_0}{r^{2(1-\alpha)}}\left(v \frac{\diff v}{\diff r} \right)^{\alpha}
\end{equation}
where tildes have been omitted. The first term is the modified gravity (\ie which includes the radiative pressure on free electrons) and $\aleph_0$ is a dimensionless quantity which engulfs the other variables. From now on, we model the motion of the flow above the sonic point, where the behaviour of the gas is well described by this ballistic equation, and amalgamate it with the stellar surface. We thus do not consider intrinsically time variable behaviours related to internal shocks or instabilities (\eg clumps) within the wind at the orbital scale such as sometimes described in the literature : see \eg \cite{Walder2014} where the shocks appear in the wake and consequently to the presence of the compact object (ahead of it the flow remains supersonic). It can be shown \citep{Lamers1999} that the value of $\aleph_0$ is not free, if we force the uniqueness of the solution. If we introduce $G$ the gravitational constant, $M_*$ the mass of the star, $\Gamma$ its Eddington factor\footnote{\ie the ratio of the stellar to the Eddington luminosity.} and $V$ and $R$, two fiducial scaling quantities (resp. for velocities and lengths), $\aleph_0$ is given by :
\begin{equation}
\label{eq:aleph0}
\aleph_0=\frac{1}{\alpha}\left( \frac{\alpha}{1-\alpha} \right)^{1-\alpha} \left[ \frac{GM_*(1-\Gamma)}{RV^2} \right]^{1-\alpha}
\end{equation}
In the case of an isolated star, one can always choose $R$ and $V$ such as the term between brackets in equation \eqref{eq:aleph0} is 1 and the equation of motion takes the form of equation \eqref{eq:CAKadim} which can be integrated to find the well-known analytical expression of the velocity profile. When the terminal speed $v_{\infty}$, typically several times the effective escape velocity $v_{\text{esc}}^2=[2GM_*(1-\Gamma)]/R_*$, turns out to be large with respect to the initial speed at the stellar surface, typically smaller than the sound speed, we have :
\begin{equation}
\label{eq:vr_ana}
v(r)=v_{\infty}\sqrt{1-\frac{R_*}{r}} \quad \text{with} \quad v_{\infty}=v_{\text{esc}}\sqrt{\frac{\alpha}{1-\alpha}}
\end{equation}
In anticipation of the numerical integration of the full problem, no longer analytically solvable, we tested the integrator, described in \ref{sec:numTools}, on equation \eqref{eq:CAKadim}. The results are presented in Figure\,\ref{fig:1DCAK} and compared to the analytical solution \eqref{eq:vr_ana}. Notice that this velocity profile is a specific case of the more phenomenological $\beta$-laws profiles fitted to the data with $v_{\infty}$ and a free exponent, $\beta$, not necessarily $1/2$ like above. The numerically derived velocity profiles and terminal velocities match the analytical ones within $10\%$, a precision good enough to investigate forward the modified \textsc{cak} model described below.

\subsubsection{Relaxing the finite cone angle assumption}		
Since the compact object lies generally within a couple of stellar radii in a \sgx and the accretion will critically depend on the velocity of the wind as it passes close to it, the initial phase of acceleration of the wind must be accurately modelled. An improvement on the theory described above in the vicinity of the launching surface is to consider the star no longer as a point source but as an extended emission region and integrate over all the incoming directions of the photons \citep{Pauldrach1986,Friend1986,Kudritzki1989a}. Accounting for the finite size of the disc translates into an additional dimensionless factor in equation \eqref{eq:CAK} given by :
\begin{equation}
\label{eq:fin_disc_D}
D(r,\diff v/\diff r;\alpha,R_*)=\frac{(1+\sigma)^{1+\alpha}-(1+\sigma \mu_\textsc{c}^2)^{1+\alpha}}{(1+\alpha)(1-\mu_\textsc{c}^2)\sigma (1+\sigma)^{\alpha}}
\end{equation}
with $\sigma=( \diff \log v / \diff \log r ) - 1 $ and $\mu_\textsc{c}^2 = 1-(R_*/r)^2$. For $r>>R_*$, one retrieves $D\sim 1$ but for $r\sim R_*$, $D$ reaches its lowest value above the sonic surface, $D_0=1/(1+\alpha)<1$, which rises the critical value of $\aleph_0$ required to make the launching possible (since otherwise, the \cak acceleration would no longer be large enough to overcome gravity). It has two effects on the issued velocity profile. First, the exponent in the expression of the velocity profile rises, in the configurations we considered up to approximately $~0.7$, making the profile smoother. It is consistent with observations \citep{Villata1992} and sophisticated semi-analytical computation of modified \textsc{cak}-wind profiles \citep{Pauldrach1986,Villata1992,Muller2008,Araya2014a}, though on the lower end of the range. Second, the terminal speeds obtained are larger. Following the suggestions of recent state-of-the-art hydrodynamical simulations \citep{Muller2008,Noebauer2015} and of analytical studies \citep[see][Table 1]{Kudritzki1989a} about the terminal speeds, we compared our results to the following velocity profile :
\begin{equation}
\label{eq:vr_FD}
v(r)=v_{\infty}\left(1-\frac{R_*}{r}\right)^{0.7} \quad \text{with} \quad v_{\infty}\sim 2.5 v_{\text{esc}} \frac{\alpha}{1-\alpha}
\end{equation}
where the factor $2.5$ holds for effective stellar temperatures beyond 21,000K \citep{Vink1999}. Notice that this law leads to generally higher terminal speeds than the one suggested by \cite{Friend1986} but to similar values than the observed ones for class I luminosity stars \citep[see table 3 of ][]{Groenewegen1989}, if one considers $\alpha\sim 0.55$. Once again, we used our numerical integrator to compute the velocity profile and compared it to the profile above in Figure\,\ref{fig:1DCAK}. The numerically derived velocity profiles and terminal velocities match the empirical values within a few percents.

\subsection{Scale invariant expression of the wind motion in a Roche potential}
\label{zemodel}
\begin{figure}
\def\svgwidth{235pt}
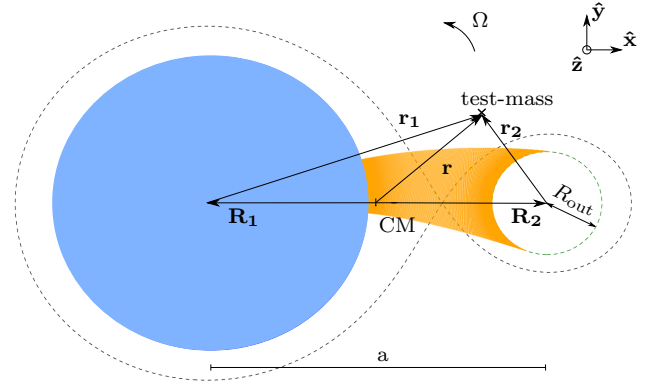
\caption{Orbital slice of the \sgx configuration we consider with, in blue, the supergiant stellar companion and on the right, the compact object. The eight-shaped black dotted line is the Roche surface passing through the first Lagrangian point and the orange lines are the computed streamlines which enters the vicinity of the compact object. The latter, in dashed green, is defined by its radius $R_{\text{out}}$ as specified in section \ref{sec:mod_acc_rad}. $a$ is the orbital separation.}
\label{fig:sketch}
\end{figure}

\sgx feature low-eccentricity orbits compared to the other classes of \textsc{hmxb} : we assume the system to be fully circularized, and the stellar spin to be synchronized \citep{Claret1997,VanEylen2016}. We now set ourselves in the frame co-rotating at the orbital period\footnote{And so do all the velocities we refer to in this article.}, with the center of mass (\textsc{cm}) at the origin and use subscript 1 (resp. 2) to refer to the star (resp. to the compact object), as described in Figure\,\ref{fig:sketch}. The modified gravitational force of the star, the gravitational force of the compact object and the centrifugal force acting on a test-mass can then be written as a potential, the Roche potential, using Kepler's third law : 
\begin{equation}
\Phi_{\textsc{r}}(\mathbf{r})=\Phi_0 \cdot \tilde{\Phi}(\mathbf{\tilde{r}};q,\Gamma)
\end{equation}
with $\Phi_0=GM_2/a$ the scale of potential, $q=M1/M2$ the mass ratio, $a$ the orbital separation, $\mathbf{r}$ the position vector of the test-mass and $\tilde{\Phi}$ thus defined by, if we use $a$ as the length scale :
\begin{equation}
\label{eq:phi_tilde}
\tilde{\Phi}(\mathbf{\tilde{r}};q,\Gamma)=-\frac{q(1-\Gamma )}{\left| \mathbf{\tilde{r}}-\mathbf{\tilde{R}_1} \right|}-\frac{1}{\left|  \mathbf{\tilde{r}}-\mathbf{\tilde{R}_2} \right|} + \frac{1}{2}(1+q) \mathbf{\tilde{r}_{\bot}}^2
\end{equation}
with $\mathbf{\tilde{r}_{\bot}}=\mathbf{\tilde{r}}-(\mathbf{\tilde{r}}\cdot\mathbf{\hat{z}})\mathbf{\hat{z}}$ where $\mathbf{\hat{z}}$ is the unit vector normal to the orbital plane and oriented in the direction of the orbital rotation vector $\mathbf{\Omega}$ (see Figure\,\ref{fig:sketch}). $\mathbf{\tilde{R}_1}$, $\mathbf{\tilde{R}_2}$ and $\mathbf{\tilde{r}}$ are the dimensionless position vectors of the star, of the compact object and of the test-mass respectively, the two first ones depending only on the mass ratio. One must also consider the Coriolis force per mass unit (which cannot be written as a potential).
At this point, the mass ratio $q$ and the stellar Eddington parameter $\Gamma$ are the only shape parameters of the problem. The radius of the star does not play any role in the dynamics since in Roche's formalism, both objects are point-like. Numerically, it means that $q$ and $\Gamma$ are the only meaningful parameters for the study of the motion of a test-mass in a Roche potential accounting for continuous radiative pressure, since the scaling back to physical units can be realized afterwhile. 

We can now gather those terms with the absorption lines acceleration described in the previous section, but so as to reduce the number of shape parameters to its very minimum and to explore the parameter space efficiently, we change the normalization variables to :

\begin{enumerate}
\item the stellar Roche lobe radius $R_{\textsc{R,}1}$ as the length scale
\item the mass of the compact object, $M_2$, as the mass scale
\item $GM_2/R_{\textsc{R,}1}^2$ as the acceleration scale
\end{enumerate}
Unless explicitly stated further in this article, this system of units is the reference we are relying on. The other scales are deduced by the simplest combinations of scales above, without any additional dimensionless factor. To consistently incorporate the \textsc{cak}-wind to the Roche potential framework, we assume that the radiatively-driven acceleration term is not altered by the presence of the compact object, which is acceptable for large mass ratio systems such as \sgx. For instance, we still assume the mass density to decrease in a spherically symmetric way around the star. Finally, let us write the approximate ratio of the radius of the Roche lobe of the star to the orbital separation, $\mathcal{E}(q)$, as suggested by \cite{Eggleton1983} :

\begin{equation}
\mathcal{E}(q)=\frac{0.49q^{2/3}}{0.6q^{2/3}+\ln{(1+q^{1/3})} }
\label{eq:egg}
\end{equation}
which leads us to adimensioned expressions of the positions of the test-mass relatively to the star and the compact object : 

\begin{equation}
\left\{
\begin{array}{ll} 
\mathbf{\tilde{r}_1}=\mathbf{\tilde{r}}-\mathbf{\tilde{R}_1}=\mathbf{\tilde{r}}+\frac{1/\mathcal{E}(q)}{q+1}\mathbf{\hat{x}}\\
\mathbf{\tilde{r}_2}=\mathbf{\tilde{r}}-\mathbf{\tilde{R}_2}=\mathbf{\tilde{r}}-\frac{q/\mathcal{E}(q)}{q+1}\mathbf{\hat{x}}
\end{array} 
\right.
\end{equation}
where $\mathbf{\hat{x}}$ is the unit vector of the axis joining the two bodies, oriented from the star to the compact object. Omitting the tildes and using the radial velocity of the test-mass relatively to the star, $v_1=\mathbf{v}\cdot\left(\mathbf{r_1}/r_1\right)$, we then get the following adimensioned ballistic equation of motion of a radiatively-driven wind in a \sgx Roche potential :
\begin{align}
\label{eq:main}
\frac{\diff \mathbf{v}}{\diff t} =& - \left[ \frac{q(1-\Gamma )}{r_1^3} 	- \aleph \cdot D\left(r_1,\frac{\diff v_1}{\diff r_1};\alpha,f\right) \frac{v_1^{\alpha}}{r_1^{3-2\alpha}} \left( \frac{\diff v_1}{\diff r_1}\right)^{\alpha}  \right] \mathbf{r_1} \nonumber\\
& - \frac{1}{r_2^3}  \mathbf{r_2} + (1+q)\mathcal{E}^2 \mathbf{r_{\bot}} -2\mathcal{E}^{3/2}\sqrt{1+q}\cdot\mathbf{\hat{z}}\wedge\mathbf{\tilde{v}}
\end{align}
with $f=R_1/R_{\text{R,}1}$ the filling factor\footnote{where $R_1$ is the stellar radius, not to be confused with the norm of $\mathbf{R_1}$.} and $\aleph=\aleph_0/D_0>\aleph_0$ the modified critical $\aleph$ to account for the finite cone angle factor at the surface. With the normalization we chose, the expression of $\aleph$ becomes :
\begin{equation}
\aleph=\frac{1+\alpha}{\alpha} \left[ \frac{\alpha}{1-\alpha}  q(1-\Gamma) \right]^{1-\alpha}
\end{equation}
and can be determined a priori from the values of $\alpha$, $q$ and $\Gamma$. The two additional shape parameters which appear in the equation \eqref{eq:main} compared to equation \eqref{eq:phi_tilde} are the $\alpha$ force multiplier and the filling factor (which plays a role through the finite disc factor but more importantly, in the way it sets the initial conditions to be verified by this equation). Those four degrees of freedom are the cornerstone of the \sgx toy-model we investigate at large scale in this article. 

\section{Numerical process}
\label{sec:second_sec}

\subsection{The modified accretion radius}
\label{sec:mod_acc_rad}
As visible in Figure\,\ref{fig:1DCAK}, the neutron star typically lies within the acceleration zone of the wind. This specific feature of \sgx compared to symbiotic binaries makes the accretion properties of the system possibly very sensitive to the ratio of the orbital separation to the characteristic acceleration length scale of the wind. One way to evaluate the amplitude of this accretion, is to compute the accretion radius from the classical Bondi-Hoyle-Lyttleton (henceforth \textsc{bhl}) framework \citep[][for a review]{Hoyle:1939fl,Bondi1944,Edgar:2004ip} : for a planar flow with a relative supersonic velocity at infinity $v_{\bullet}$ with respect to a body of mass $M_2$, the accretion radius is given by $R_{acc}=(2GM_2)/v_{\bullet}^2$. It is the critical impact parameter below which matter is likely to be accreted. Concerning $v_{\bullet}$, we get it by computing the velocity reached by a test-mass once it reaches the position of the compact object, but without the gravitational term of the neutron star in equation \eqref{eq:main}. Given the typical values of $v_{\bullet}$ ($\sim 1000$\,km$\cdot$\,s$^{-1}$), the flow is expected to be supersonic from the sonic surface - assimilated to the stellar one in our model, as suggested by equation (8) in \cite{Ducci2009} - to the vicinity of the compact object. Numerical simulations of \textsc{bhl} accretion show that the flow remains supersonic around the accretor beyond a few $R_{acc}$ \citep{ElMellah2015}. We then define an extended accretion sphere of radius $R_{\text{out}}=8R_{acc}$ around the accretor where we stop our ballistic simulations. Within this sphere, a proper hydrodynamical treatment is compulsory, in particular in the wake of the accretor, and will be carried on in an upcoming paper. From now on, we refer to $R_{\text{out}}$ and the sphere it defines as "extended" accretion radius and sphere.

\subsection{Numerical implementation}
\label{sec:numTools}
To solve the equation of motion \eqref{eq:main}, we applied a fourth order Runge-Kutta scheme on test-masses starting at the surface of the star\footnote{By symmetry, only the upper half of the system has been considered.} with a zero velocity. Notice that we do not consider the departure from the spherical shape of the stellar surface nor the induced inhomogeneities of temperature. To initiate the wind and compute the radial velocity gradient in equations \eqref{eq:CAK} and \eqref{eq:fin_disc_D}, we force the first step with radial and velocity steps small compared to the stellar radius and to the velocities computed afterwards. We validated the invariance of our code to changes in those two fiducial steps. The time steps are chosen to be smaller by a factor of 100 than any time-homogeneous combination of the displacement, the velocity and the acceleration so as to not alter the outcome of the numerical integration. The time steps are shortened again as the test-mass approaches the extended accretion sphere. Once the integration brought the test-mass trajectory to intercept (resp. miss) the extended accretion sphere around the compact object, its position and velocity on the sphere is saved (resp. discarded). To better resolve the borders and improve the statistical significance of our computation of the characteristic quantities of the inflow within the extended accretion sphere, we then refine the initial positions at the surface of the star which were selected and repeat the integration until we are left with a few 10,000 selected arrival points to process. Some of the streamlines susceptible to be accreted have been represented in orange in Figure\,\ref{fig:sketch} in a fiducial case - a similar sketch for each configuration $(q,f,\Gamma,\alpha)$ is automatically produced. The whole computation time per simulation on 1 \textsc{cpu}, including the post-processing described in the next paragraph, is of the order of a few hours such as the adimensioned parameter space considered in section \ref{sec:param_visu} can be explored within a week with 36 \textsc{cpu}s. The empirical validation of the integrator has been carried on in the previous section on the wind of isolated hot stars.

Concerning the primitive quantities at the extended accretion sphere, we save the velocity and positions of the test-masses when they reach it. Since our model is purely ballistic, we cannot access the density directly but we can get the divergence of the streamlines by monitoring the evolution of the surface elements delimited by the four closest streamlines at the departure and at the arrival. Once a physical scale has been chosen, we can then deduce from this value the density corresponding to each streamline on the extended accretion sphere. An angular resampling of those primitive quantities (mass density and velocity) on pre-defined meshes provides us with physically-motivated outer boundary conditions for upcoming hydrodynamical simulations on the fate of this inflow within the extended accretion sphere.


\subsection{Parameters and visualization}
\label{sec:param_visu}

We sample the four shape parameters accordingly to the values expected in a neutron star hosting \sgx :
\begin{enumerate}
\item $\alpha$ is either 0.45, 0.50, 0.55 or 0.65. If the latter value is more suited for early supergiant O type stars, the three first ones correspond to the values calculated by \cite{Shimada1994} from 520,000 atomic lines for early B / late O supergiant stars. The value of $\alpha$ depends mostly on the effective stellar temperature $T$ and the decimal logarithm of the surface gravity, $\log (g)$. 
\item $\Gamma$, the Eddington parameter. For OB-supergiants, we expect it to be below 30\%, by opposition to Wolf-Rayet stars, Luminous Blue Variables or hypergiant stars (like the donor in GX 301-2) which can go beyond. We take 10, 20 and 30\% as the three possible values.
\item $q$, the mass ratio. We ran simulations for 12 integer values of $q$ ranging between 7 and 18.
\item $f$, the filling factor. We picked up 10 values non regularly spaced between 50\% and 99\%.
\end{enumerate} 
To access physical outputs (such as the stellar radius or the mass accretion rate onto the compact object), we also need user-specified scale parameters : the orbital period, the mass of the compact object and the $Q$ force multiplier (to compute the physical mass loss rate of the star). According to \cite{Shimada1994}, $Q$ lies in a narrow range around 900 for early-type B supergiant stars with effective temperatures above the bi-stability threshold \citep[$\sim$20 kK,][]{Vink1999} and below 30 kK.

Using the Spyre library for Python developed by A. Hajari (2015), we designed a handy web application framework to quickly visualize all the configurations in the scope of this article. It displays colormaps in the $(f,q)$ plane for many different outputs, computed based on the simulations results (\ie the shape parameters) and, for the dimensioned outputs, on the user-specified scales. Those outputs have been chosen due to their relevance to appreciate the properties of the model we developed and the interplay between the different variables. This interface, available on demand, aims at conducting the user towards a better understanding of the trends we observe in \sgx by quickly travelling through a reduced though essential parameters space. It also entitles the user to evaluate the dependencies of each output with the 7 shape and scale parameters and, where suggested by the figures, to neglect some of those dependencies or emphasize the degeneracies between parameters.

\section{Structure and properties of the flow at the orbital scale}
\label{sec:structure_wind}


\subsection{Structure of the accretion flow}
\label{sec:struct_acc_flow}
We now try to quantify what we mean by "wind" accretion. The usual smoking gun to qualify an accretion mechanism of wind accretion in a binary system is to observe an underflowing star with an accreting companion. However, it does not tell us much about the actual structure of the flow. Indeed, a wind which would not be fast enough compared to the orbital velocity once it reaches the critical Roche surface would result in a highly collimated flow, mimicking the stream one usually associates to \rlof \citep{Nagae2005,Mohamed}. \rlof tells us something about the filling factor while wind accretion refers to the structure of the flow, not necessarily incompatible with \rlof. Thus, we choose to term "wind" or "stream" an accretion flow depending on the relative size of the extended accretion sphere, where the shock is expected to develop, with respect to the size of the Roche lobe of the compact object $R_{\textsc{r,}2}$. When the wind is slow, the former becomes of the order or even larger than the latter and the whole orbital scale requires a proper hydrodynamical treatment. 

\begin{figure}
\begin{center}
\includegraphics[width=\columnwidth]{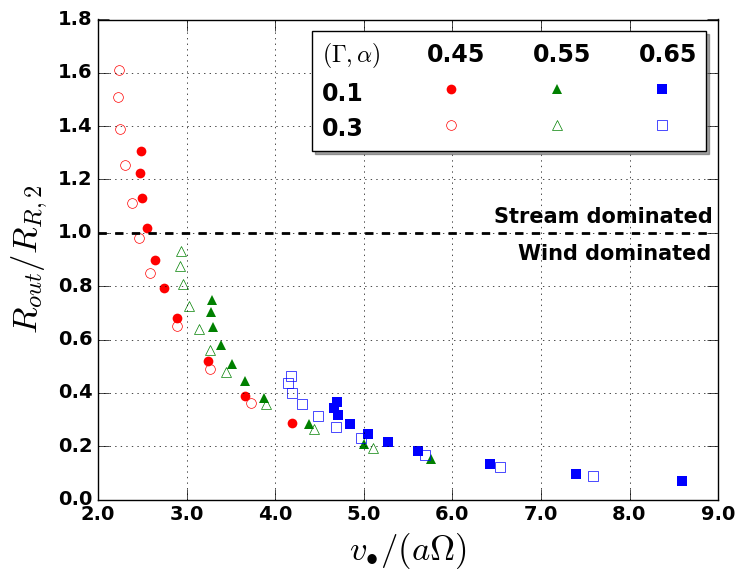}	
\caption{Ratio of the extended accretion radius to the size of the compact object Roche lobe as a function of the ratio of the velocity at the orbital separation to the orbital speed. For each of the six curves, from upper left to lower right, we used values of $(f,q)$ along the line going from $(0.99,8)$ to $(0.5,17)$ in Figure\,\ref{fig:beta_beta_HL} so as to probe the largest range of $R_{\text{out}}/R_{\text{R,}2}$.}
\label{fig:RLOF_VS_wind}
\end{center}
\end{figure}

To evaluate how slow the wind must be to be in such a configuration, we plotted in Figure\,\ref{fig:RLOF_VS_wind} the evolution of the ratio of the aforementioned length scales compared to the ratio of $v_{\bullet}$ to the orbital speed. Those adimensioned outputs do not require the specification of any scale and depend only on the shape parameters. Besides, none of these quantities was forced ; instead, they were computed from parameters linked to the launching of the wind and the structure of the potential. A first conclusion is that, even in configurations where the filling factor is close to one, a high velocity flow (a wind) is expected for efficient wind acceleration (\ie $\alpha > 0.55$), for any Eddington factor below 30\%. It is a typical situation where \rlof cohabits with wind accretion. On the other hand, even for a filling factor as low as 90\%, one can get a stream dominated flow for $\alpha=0.45$, intermediate mass ratios and Eddington factors of 30\%. We observe that lower values of $\Gamma$, \ie of the adimensioned luminosity, favor a wind dominated flow, which can sound paradoxical but is actually consistent with the velocity of the wind scaling as the modified escape speed in equation \eqref{eq:vr_FD}.

An important conclusion to draw from this Figure\,\ref{fig:RLOF_VS_wind} is also that, as far as the structure of the accretion flow is concerned, the ratio $v_{\bullet}/a\Omega$ is an excellent tracer to evaluate how windy the flow looks, much more reliable than the filling factor which only plays a secondary role here. In the case of radiatively-driven winds, we also advocate in favour of this ratio rather than the ratio of the mean normal velocity of the wind on the critical Roche surface to the orbital speed, $V_R/a\Omega$, as suggested in \cite{Nagae2005} and \cite{Jahanara2005}. Indeed, first, since the accretor lies within the wind acceleration radius, it is more relevant to consider the velocity of the wind at the orbital separation than at the critical Roche surface which must not be considered as seriously in \sgx as in systems hosting low $\Gamma$ stars (symbiotic binaries, \textsc{lmxb}...) : in Figure\,\ref{fig:sketch}, the critical Roche surface is represented for $\Gamma=0$ but in \sgx, the lowering of the effective stellar gravity by the radiative pressure from the continuum opacity is no longer negligible \citep{Dermine2009}. Second, it is not equivalent to model the wind as described in section \ref{sec:windModel} and to consider that the radiative pressure exactly balances the stellar gravitational force at any radius (ie $\Gamma=1$ and $g_{\textsc{cak}}=0$), as used in \cite{Theuns1996d} and \cite{Jahanara2005}. It is legitimate in symbiotic binaries where the mechanism responsible for the driving of the wind is still poorly known but not in \sgx. As a consequence, the conservative threshold on $v_{\bullet}/a\Omega$ we found, of the order of 2 to 3, should prove to be more reliable in \sgx than $V_R/a\Omega\sim 0.5$ derived in \cite{Nagae2005}. 

\subsection{Fraction of wind captured}
\label{sec:mass_acc_rate}

\subsubsection{Direct numerical measures}

First, we save $\beta ^+$, the fraction of stellar wind entering the extended accretion sphere. The "$+$" superscript refers to the fact that not all the matter entering the extended accretion sphere is expected to be accreted - but all the finally accreted streamlines belong to this sample ; the fraction of stellar wind entering the extended accretion sphere is an upper limit on the fraction of stellar wind being accreted, $\beta$. A first underestimation of $\beta ^+$ is obtained by comparing the number $n$ of streamlines which were selected (in the sense defined in section \ref{sec:numTools}) to the total number $N$ of streamlines within the launching patchwork considered\footnote{A subsample of the total stellar surface which contains all the streamlines which will be selected.} and then by extending it to account for the whole sphere :
\begin{equation}
\label{eq:betaplus_first}
\beta ^+ = \frac{n}{N} \times \frac{\frac{\pi}{2}\times \sin\left[ \arctan\left( \frac{2R_{\text{out}}}{a} \right) \right]}{2\pi}
\end{equation}
where the numerator of the second factor of the \rhs is the solid angle covered by the launching patchwork. $\beta ^+$ is an underestimation because it does not account for the fact that the classical meshing of the sphere we rely on is not homogeneous. It is less dense at the equator, from where come most of the streamlines selected, than at the poles. Another way to evaluate $\beta ^+$ is to compute the surface delimited on the stellar sphere (in blue in Figure \ref{fig:sketch}) by the streamlines which enter the extended accretion sphere. To do so, we compute the mean arc $\left\langle l \right\rangle$ between selected streamlines and compute :
\begin{equation}
\beta ^+ = \frac{n \times \left\langle l \right\rangle^2}{2\pi}
\end{equation} 
In practice, equation \eqref{eq:betaplus_first} is only a slight underestimation of the order of a few percents compared to the second estimation of $\beta ^+$. 

Concerning the actual fraction of the wind which will take part in the feeding of the compact object and in the subsequent emission of X-rays, we can not make any definitive measure without three dimensional hydrodynamical simulations of the evolution of the flow within the extended accretion sphere. However, an indirect measure of $\beta$ can be obtained using the results of \cite{ElMellah2015}. Indeed, if we write $\dot{M}^+_{\textsc{ec15}}$ the mass inflow which was arbitrarily set as an outer boundary condition in \cite{ElMellah2015} and $\dot{M}_{\text{acc}}$ the mass accretion rate which was then measured for Mach numbers at infinity larger than 4, the fraction of the wind entering the sphere of radius $8R_{acc}$ which was actually accreted is given by :
\begin{equation}
\label{eq:frac_in}
\frac{\dot{M}_{\text{acc}}}{\dot{M}^+_{\textsc{ec15}}}=\frac{0.77\dot{M}_{\textsc{hl}}}{\rho_{\infty}v_{\infty}\pi\left(8R_{acc}\right)^2}=\frac{0.77}{8^2}\sim 1\% \sim \frac{\beta}{\beta^+}
\end{equation}
where $\rho_{\infty}$ and $v_{\infty}$ are the mass density and the velocity at infinity\footnote{Approximation acceptable since the flow is almost not disturbed from planarity when it enters the simulation space at $8R_{acc}$ - see outer boundary conditions (9) to (11) in \cite{ElMellah2015}.} and $\dot{M}_{\textsc{hl}}$ is the mass accretion rate according to the \bhl approach (see section \ref{sec:bhl_estim} below). Remarkably enough, this ratio \citep[whose independence with the numerical setup and the size of the compact accretor has been shown in][]{ElMellah2015}, no longer depend on any parameter and remains fairly constant when the Mach number is above 4, which is easily verified by the highly supersonic winds in \sgx.

\subsubsection{\textsc{bhl} estimation of the accreted wind fraction}
\label{sec:bhl_estim}

To have a theoretical expectation to compare to, we can use the computed adimensioned velocity of the wind at the distance of the orbital separation, $v_{\bullet}$, to make predictions on the fraction of the wind being accreted by the compact object according to a \textsc{bhl} approach \citep{Boffin1988}. Corrected with an additional factor $0.77$, it has been shown to be accurate within a few percents for Mach numbers at infinity larger than a few \citep{Foglizzo1996,ElMellah2015}. If we assume the wind to be isotropic, we have, using the conservation of mass :
\begin{equation}
R_1^2 \rho _1 v_1 \sim a^2 \rho _{\bullet} v_{\bullet}
\end{equation}
with the $\bullet$ subscript referring to the values at a distance from the stellar center equal to the orbital separation $a$ and the subscript 1, to the values at the stellar surface. Then, the fraction of the wind accreted is given by :
\begin{equation}
\label{eq:betaHL}
\beta_{\textsc{hl}} \hat{=} \frac{\dot{M}_{\textsc{hl}}}{\dot{M}_1} = \frac{0.77\cdot \pi R_{acc}^2 \rho_{\bullet}v_{\bullet}}{4\pi R_1^2 \rho _1 v_1} \sim \frac{0.77}{4}\left( \frac{R_{acc}}{a} \right)^2 \sim 0.77\times\frac{\mathcal{E}^2(q)}{v_{\bullet}^4}
\end{equation}
where $\dot{M}_1$ and $\dot{M}_{\textsc{hl}}$ are respectively the stellar mass outflow and the \textsc{bhl} mass accretion rate. A decisive point we emphasized in equation \eqref{eq:betaHL} above is the independence of $\beta_{\textsc{hl}}$ on the scaling : whatever the period, the orbital separation or the mass of the compact object, the fraction of the wind accreted according to Bondi, Hoyle \& Lyttleton's sketch is the same if the four shape parameters remain unchanged. This comment will be of prime importance in the interpretation of the results presented in section \ref{sec:phys_mass_acc_rate}.

\subsubsection{Assessment of the theoretical tracers}

\begin{figure}
\begin{subfigure}{.5\textwidth}
\centering
\includegraphics[width=0.95\columnwidth]{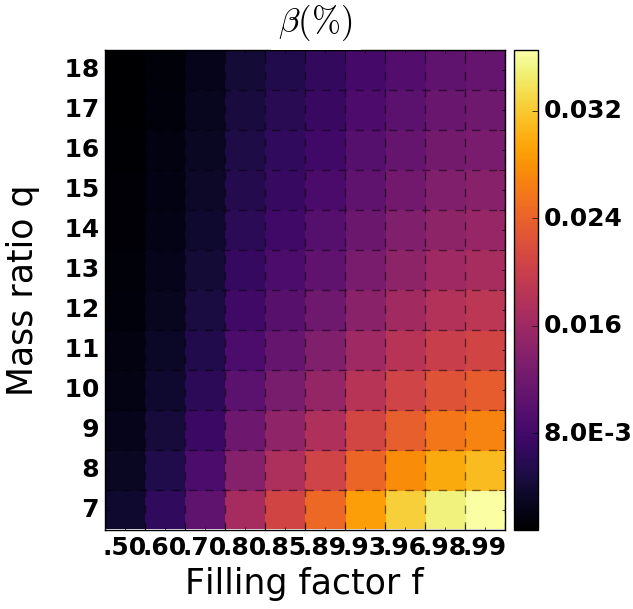}
  \label{fig:sfig1}
\end{subfigure}
\begin{subfigure}{.5\textwidth}
\centering
\includegraphics[width=0.95\columnwidth]{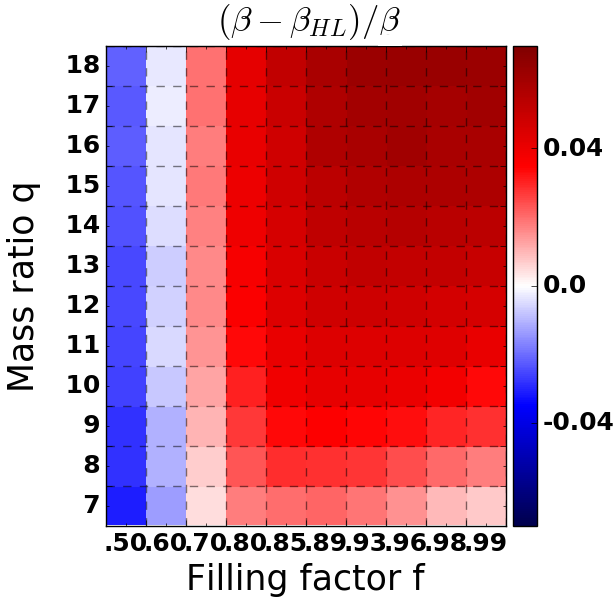}
  \label{fig:sfig2}
\end{subfigure}
\caption{(upper panel) Measured fraction of wind accreted for $\alpha=0.45$ and $\Gamma=0.2$ as a function of the filling factor and the mass ratio. (bottom panel) Relative discrepancy between the measured fraction of wind accreted and the one predicted by the \bhl prescription \eqref{eq:betaHL} used with the computed velocity at the orbital separation $v_{\bullet}$.}
\label{fig:beta_beta_HL}
\end{figure}

The relevance of the wind-\rlof model for symbiotic binaries is a suggestive hint that wind accretion can prove more efficient than what the \textsc{bhl} prescription would suggest. We represent in Figure\,\ref{fig:beta_beta_HL} the measured fraction of wind accreted $\beta$ in the upper panel and the departure from the fraction derived from a \bhl approach, $\beta_{\textsc{hl}}$. The bottom part of Figure\,\ref{fig:beta_beta_HL} indicates that the enhancement of the fraction of wind being accreted with respect to the \bhl prescription is low in \sgx, weaker than in symbiotic binaries \citep{Mohamed}. Provided the wind speed at the orbital separation $v_{\bullet}$ (and not the terminal speed $v_{\infty}>v_{\bullet}$) is used to compute $\beta_{\textsc{hl}}$, we retrieve the measured value of $\beta$ within 6\% for the mass ratios and the filling factors we consider. Using the terminal speed suggested in formula \eqref{eq:vr_FD} instead would have led to the following analytic expression of the fraction of wind being accreted :
\begin{equation}
\beta_{\infty} = 0.64\% \times \underbrace{\left(\frac{\mathcal{E}}{q}\right)^2}_{0.1\%\rightarrow 0.6\%} \underbrace{\left[\frac{f}{1-\Gamma}\cdot \left(\frac{1-\alpha}{\alpha}\right)^2\right]^2}_{<4.5}
\end{equation}
where the estimations below the underbraces are for the range of shape parameters we consider. It clearly leads to underestimated values of the fraction of wind accreted by at least a factor of three. The terminal wind speed must thus not be used to estimate the fraction of wind captured by the compact object. 

For $\beta$, it is no surprise that, at a given mass ratio, as the star fills a lower fraction of its Roche lobe, the relative distance between the stellar surface and the compact object rises. Since the terminal velocity does not depend on the stellar radius and since the stellar gravity is largely overrun by the absorption lines acceleration close to the surface, we can expect the velocity at the compact object position with respect to the unchanged orbital velocity to be larger and thus, a lower extended accretion radius $R_{out}$ compared to the unchanged orbital separation. In other words, the angular size of the extended accretion sphere as seen from the star drops and so does the fraction of the wind being accreted. However, if one considers, at a given filling factor, an increase in the mass ratio, let us say at a given mass of the compact object, things get trickier. Indeed, on one hand the terminal speed is proportional to the escape velocity which rises and on the other hand, relative distance between the stellar surface and the compact object drops, giving the wind less room to accelerate. Thus, it is not possible to conclude a priori about the evolution of $\beta$ with $q$ and numerical computation was indeed required to obtain this tendency : for any filling factor, rising the mass ratio from 7 to 18 leads to a drop in the fraction of wind accreted by a factor 3. This didactic yet insightful reasoning illustrates how misleading qualitative approaches can be in systems where variables are as entangled with each other as \sgx. 

Concerning the dependency of the fraction of wind being accreted by the compact object with respect to $\alpha$, it will be discussed in more details in the discussion devoted to the predicted X-ray luminosities (section \ref{sec:phys_mass_acc_rate}) but a first observation we make is that it follows the dependency of $R_{\text{out}}/R_{\text{R,}2}$ in Figure\,\ref{fig:RLOF_VS_wind} : as $\alpha$ rises, the wind acceleration gets more efficient and the terminal speed is higher, making the speed at the orbital separation larger and the computed accretion radius smaller. Thus, both $\beta$ and $\beta_{\textsc{hl}}$ drop by approximatively an order of magnitude as $\alpha$ rises from 0.45 to 0.65. Since a rise in $\alpha$, all other things being equal, also means a lower stellar mass outflow according to equation \eqref{eq:Mdot1}, we expect much lower X-ray luminosities for $\alpha=0.65$ than for $\alpha=0.45$, statement to be confirmed by Figure\,\ref{fig:array_LXHL} afterwards. Although the fraction of wind being accreted depends on $\alpha$, it is worth noticing that it does not depend on the $Q$ force multiplier : whatever its value, the velocity profile remains unchanged. Eventually, for the influence of $\Gamma$, a rise from 10\% to 30\% only results in a rise of a factor less than 2 for $\beta^+$ and $\beta_{\textsc{hl}}$, but it will play a more important role once its influence on the mass outflow is taken into account.

\subsection{Physical mass accretion rate and X-ray luminosity}
\label{sec:phys_mass_acc_rate}

\begin{figure*}
\centering
\includegraphics[width=2\columnwidth]{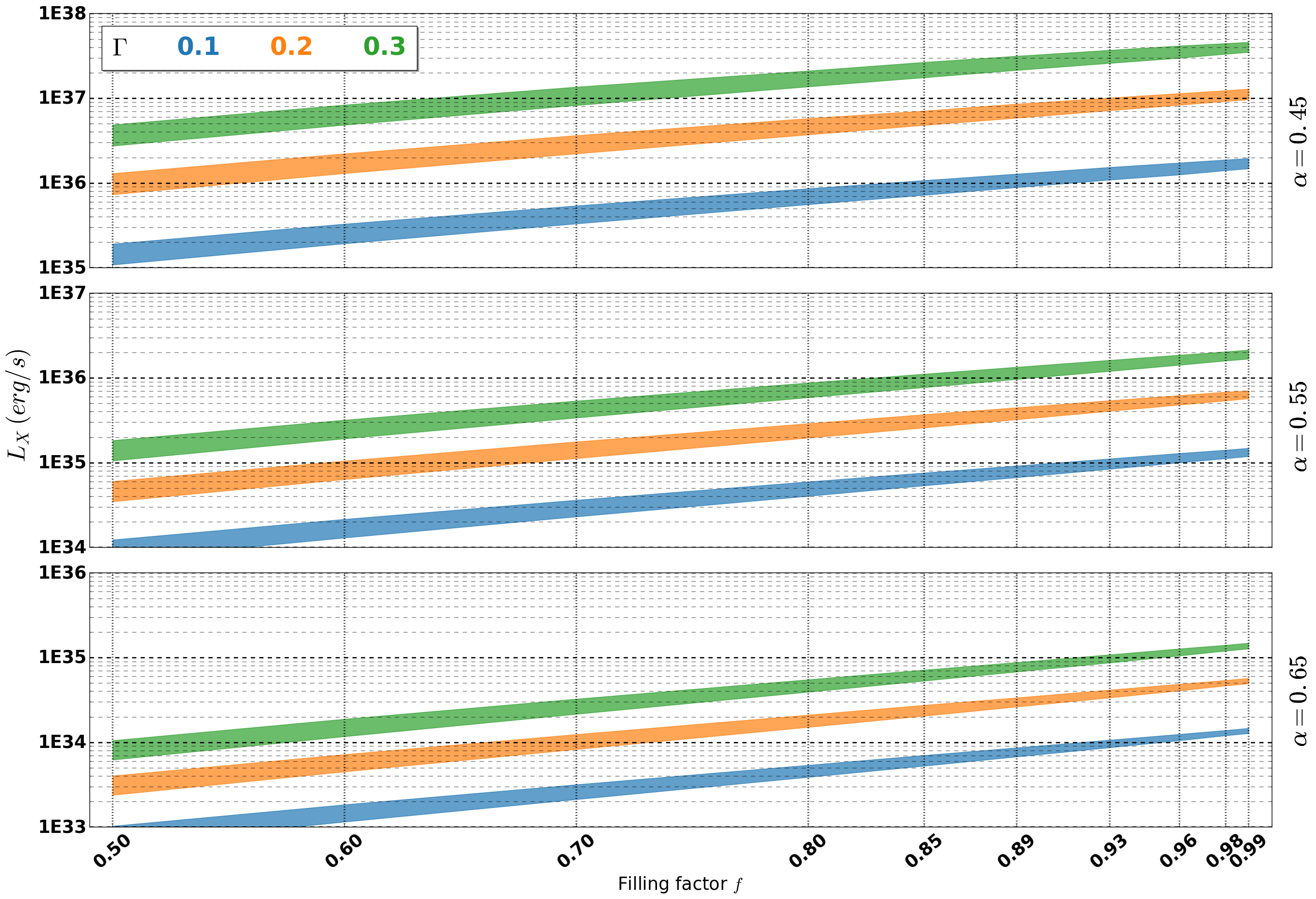}
\caption{Representation of the permanent X-ray luminosity deduced from the \textsc{bhl}'s approach (with a 25\% conversion factor) as a function of the filling factor $f$, for different $\alpha$-force multipliers and Eddington factors $\Gamma$. The weaker dependence on the mass ratio, which ranges from 7 to 18, is represented by the thickness of each line (the lower limit being for larger mass ratios) : since $M_2$ is set to 1.8\msun here, $M_1$ spans a realistic interval of 13 to 32\msun. We numerically computed the accretion radius as described in section \ref{sec:mod_acc_rad} and sampled the filling factors denoted at the bottom. Those results are computed for an intermediate value of the $Q$-force multiplier of 900. $L_X \propto Q^{(1-\alpha)/\alpha}$ according to equation \eqref{eq:Mdot1} and $Q$ is expected to range between 800 and 2,000. Changing the orbital period while keeping those quantities invariant does not change those luminosity profiles.}
\label{fig:array_LXHL}
\end{figure*}

We now go on by analyzing a scale-dependent decisive output, the physical mass accretion rate $\dot{M}_{\text{acc}}$ and the associated X-ray luminosity $L_X$. Interestingly enough, the upper panel in Figure\,\ref{fig:beta_beta_HL} shows that, for a fixed mass of the compact object, the fraction of wind accreted decreases when the mass of the supergiant star rises. However, notice that this statement, along with all the ones made in the previous section, concerns only the fraction of wind which is accreted, not the actual physical mass accretion rate ; since a larger stellar mass also means a larger absolute mass loss rate, the evolution of the X-ray luminosity with $q$ could still be different. To compute the physical mass accretion rate, we first need an estimation of the stellar mass loss rate $\dot{M}_1$. Once we account for the finite cone angle effect (first factor below), $\dot{M}_1$ writes, with the $Q$ parametrization inherited from \citep{Gayley1995} :
\begin{equation}
\label{eq:Mdot1}
\dot{M}_1=\left( \frac{1}{1+\alpha} \right)^{\frac{1}{\alpha}} \times \frac{\alpha}{1-\alpha}\Gamma\left( \frac{\Gamma Q}{1-\Gamma}\right)^{\frac{1-\alpha}{\alpha}} \frac{L_{\text{Edd}}}{c^2}
\end{equation}
where $L_{\text{Edd}}$ is the Eddington luminosity of the star (which depends only on its mass). On the other hand, we use equation \eqref{eq:frac_in} to deduce from the measured fraction of wind entering the extended accretion sphere, the total fraction of the wind which is actually likely to be accreted by the compact object. The product of $\beta$ by $\dot{M}_1$ yields the physical mass accretion rate and the corresponding X-ray luminosity is then obtained for an accretor of compactness parameter representative of a neutron star ($\sim$25\%). 

In Figure\,\ref{fig:array_LXHL} has been represented the evolution of the steady-state X-ray luminosity as a function of the 4 shape parameters. Given the restricted range of expected values for $Q$ and $M_2$ and the weak dependency of $L_X$ (\ie of $\dot{M}_1$) on them, we chose to highlight the dependence of the X-ray luminosity on the dimensionless parameters. 

The stronger dependence of $L_X$ is on the $\alpha$ force multiplier, with a luminosity divided by 20 each time alpha rises from 0.45 to 0.55 and then 0.65. Each decrescendo increment on $\Gamma$ entails a division of the X-ray luminosity by a factor of approximately 6 ; a lower effective gravity makes the wind terminal speed - which scales as the effective escape speed at the stellar surface - decrease, which leads to enhanced accretion of matter. We also notice that larger values of the $\alpha$ force multiplier lower the relative dynamical range of X-ray luminosities : fast winds let fewer room to the influence of the other parameters. The luminosity rises by about an order of magnitude from a filling factor of 50\% to a configuration where the star is close to fill its Roche lobe ($f>95$\%). The weakest dependence is on the mass ratio. It turns out that, at a fixed mass of the accretor, larger stellar masses lead to lower X-ray luminosities : in the balance aforementioned, the fact that a larger stellar mass implies a larger wind speed and thus a smaller accretion cross-section prevails over the larger mass loss rate. However, the balance between those two effects weakens significantly the dependence of the X-ray luminosity on the mass ratio, weaker than the one on the filling factor $f$, the Eddington parameter $\Gamma$ and the $\alpha$ force multiplier. Given the low luminosity levels observed for $\alpha=0.65$ and even for $\alpha=0.55$, we can affirm that most of the \sgx we observe verify $\alpha\in\left[0.45;0.55\right]$. If it matches theoretical predictions from stellar atmosphere models for early-type B stars\footnote{For solar metallicity levels, \citep{Shimada1994} gives $\alpha\in\left[0.47;0.52\right]$ for a stellar effective temperature between 20 and 30,000K.} \citep{Shimada1994}, it is a bit below the values expected for O stars ($\alpha\sim 0.6$), which might suggest that the supergiant companions in \sgx are anomalously inefficient at accelerating their winds given their spectral properties, maybe because of an anomalously low metallicity\footnote{The observational bias introduced by our limited sensibility does anyway artificially enhance the fraction of marginally low metallicity observed donors for radiatively-driven winds.}.

This Figure also enables us to pinpoint the degeneracies between the parameters. Indeed, we notice that a luminous star with a wind opaque enough\footnote{With respect to the metal absorption lines.} to be efficiently accelerated (\ie $\Gamma=0.3$ and $\alpha=0.55$) yields very similar X-ray luminosities, whatever the mass ratio or the filling factor, as a lower luminosity star with a less efficient radiatively driven acceleration (\ie $\Gamma=0.1$ and $\alpha=0.45$). 

A more spectacular property of the mass accretion rate emphasized by this parametrization (with both shape and scale parameters) is the independence, all other parameters being equal, of the X-ray luminosity on the orbital period. The key point is that, if the orbital period rises, $\beta_{\textsc{hl}}$ is not altered since the accretion radius $R_{acc}$ rises in the same proportions as the orbital separation $a$. Indeed, at a fixed filling factor, it induces a larger stellar radius and alters the modified escape speed at the stellar surface in the same proportions as the reference velocity used as a velocity scale, $\sqrt{\left(GM_2\right)/R_{R,1}}$. Because the velocity of the wind simply scales with the modified escape speed at the stellar surface, the adimensioned velocity at a distance of an orbital separation, $\tilde{v}_{\bullet}$, remains the same and so does the fraction of the wind being accreted according to equation \eqref{eq:betaHL}. On a more straightforward hand, changing the orbital period does not modify the stellar mass outflow\footnote{At least for large mass ratios for which we supposed that the mass of the neutron star was too low compared to the stellar mass to alter the launching of the wind, essentially determined by the conditions at the sonic (here, stellar) surface.}. In the end, we are left with strictly the same X-ray luminosity. As a consequence, we claim that any dependency of the observed X-ray luminosities on the orbital periods must be attributed either to a departure from the present framework or to an underneath correlation not taken into account, for instance between the Eddington factor $\Gamma$ and the stellar radius, expected to be smaller for shorter period systems not undergoing \rlof : for stellar evolution reasons, the orbital separation is likely to be negatively correlated to the filling factor. In no case the orbital period could be considered as the main culprit for the X-ray luminosity, merely as a correlated quantity whose causal relation to the permanent X-ray luminosity must be traced back.


\subsection{Shearing of the accretion flow}
\label{sec:shearing}

\begin{figure}
\begin{subfigure}{.5\textwidth}
\centering
\includegraphics[width=0.9\columnwidth]{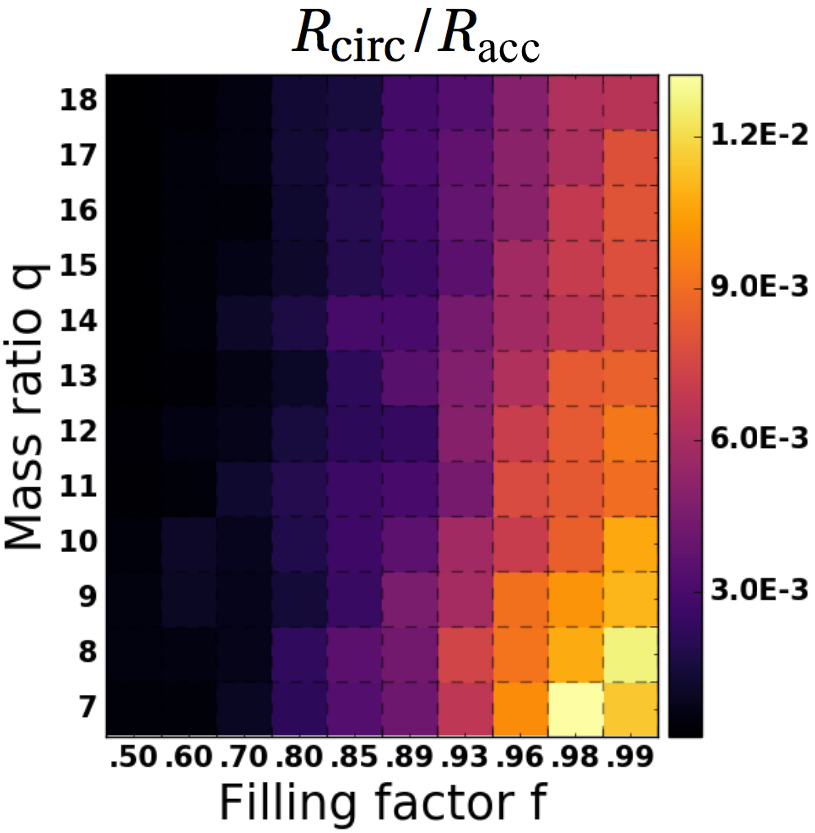}
  \label{fig:sfig1}
\end{subfigure}
\begin{subfigure}{.5\textwidth}
\centering
\includegraphics[width=0.9\columnwidth]{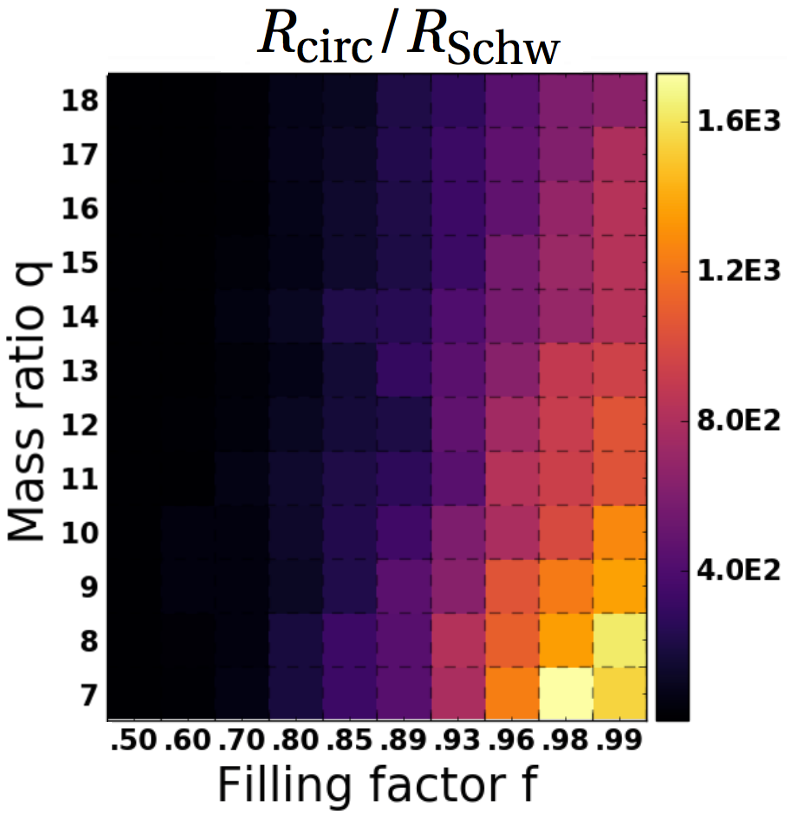}
  \label{fig:sfig2}
\end{subfigure}
\caption{(upper panel) Colormap of the relative size of the circularisation radius of the flow entering the extended accretion sphere compared to the accretion radius for $\alpha=0.55$ and $\Gamma=0.2$. (lower panel) Colormap of the relative size of the circularisation radius of the flow entering the extended accretion sphere compared to the Schwarzschild radius of the accretor for $\alpha=0.55$, $\Gamma=0.2$, $P=9$ days and $M_2=2$\msun.}
\label{fig:acc_circi_schw}
\end{figure} 

The question of the angular momentum accretion rate within the extended accretion sphere and of its dependencies on the four shape and three scale parameters remains. Do the configurations favourable to large X-ray luminosities correspond to the ones likely to give birth to a disk around the accretor?

To possibly form a disc-like structure around the accretor, the flow needs to gain enough angular momentum as it approaches the compact object. We compute an adimensioned circularization radius given the amount of angular momentum (and of mass) entering the extended accretion sphere per second. The circularisation radius can be compared to the accretion radius to shed light on the way a disk-like structure could possibly form, provided it does. The ratios $R_{acc}/R_{\text{R,}2}$ and $R_{\text{circ}}/R_{\text{R,}2}$ correlates positively to $L_X$, but more interestingly, the ratio $R_{\text{circ}}/R_{acc}$ also follows this trend, indicating that the bending of the shocked structure is stronger for low $q$, high $f$, low $\alpha$ and high $\Gamma$. In the upper panel of Figure\,\ref{fig:acc_circi_schw} has been represented this ratio as a function of $q$ and $f$ for $\alpha=0.55$ and $\Gamma=0.20$. Yet, the circularisation radius still remains two orders of magnitude smaller than the accretion radius, characteristic of the size of the shocked region ; for very inefficient wind acceleration with $\alpha=0.45$, the circularisation radius is still smaller than the accretion radius but by one order of magnitude only. We can then conclude that, apart in case of significant gain of angular momentum, a winding up of the whole shocked tail is not to be expected. If a disk forms, it grows within the shocked region \citep[as witnessed in numerical simulations by][]{Blondin2013a} where substantial instabilities are believed to take place \citep{Foglizzo2000}. Whether those instabilities drive variations of the front shock position large enough to bring the Coriolis bending in action remains to be investigated. 

The previous comments would turn out to be of little interest if the inflow crashes on the neutron star surface before it can reach its circularisation radius. Indeed, a major difference with symbiotic binaries is that in \sgx, the accretor is small enough to let plenty of room for the shock to develop and possibly, within it, to a disk to form. Once a scale is set, the ratio $R_{\text{circ}}/R_{\text{Schw}}$ can be represented, along with $R_{\text{circ}}/R_{acc}$, for intermediate values of the $\alpha$ and $\Gamma$ parameters (bottom panel in Figure\,\ref{fig:acc_circi_schw}). It shows that the circularization radius lies approximately two orders of magnitude below the accretion radius and two to three orders of magnitude above the neutron star surface. For the values of $\alpha$ and $\Gamma$ considered above and the fiducial parameters $q=13$ and $f=89$\%, the accretion radius is approximately one hundredth of the orbital separation ; according to Figure\,\ref{fig:acc_circi_schw}, the circularization radius is approximately 500 times smaller than this accretion radius and 200 times larger than the Schwarzschild radius of the neutron star (so approximately 100 times larger than the neutron star itself for a compactness parameter of $\sim 25\%$). Consequently, we retrieve the ratio 100$\times$500$\times$100$=$5$\cdot$10$^6$ between the orbital separation and the characteristic size of the neutron star in a representative \sgx. This intermediate position of the circularisation radius between the shock and the neutron star surface tends to immunize a putative disk both against disruptive instabilities at the shock level \citep{Manousakis2015c} and against premature truncation by the magnetosphere, provided the angular momentum does not significantly evolve within the extended accretion sphere. More importantly, we notice that for $\alpha$ at $0.55$ or below and for moderately high values of the filling factor, the likelihood to form a disk follows the same trend as the one drawn by the X-ray luminosity : the most luminous systems are also more likely to feature a disk-like structure around the accretor.


\section{Wind accretion on stage}
\label{sec:dim_res}

We now go on by assessing the self-consistency of our toy-model ; how far from the observations can be the plethora of parameters we have access to from a reduced number of key quantities? 


\subsection{A roadmap to self-consistent system parameters}

\begin{figure*}
\centering
\includegraphics[width=2\columnwidth]{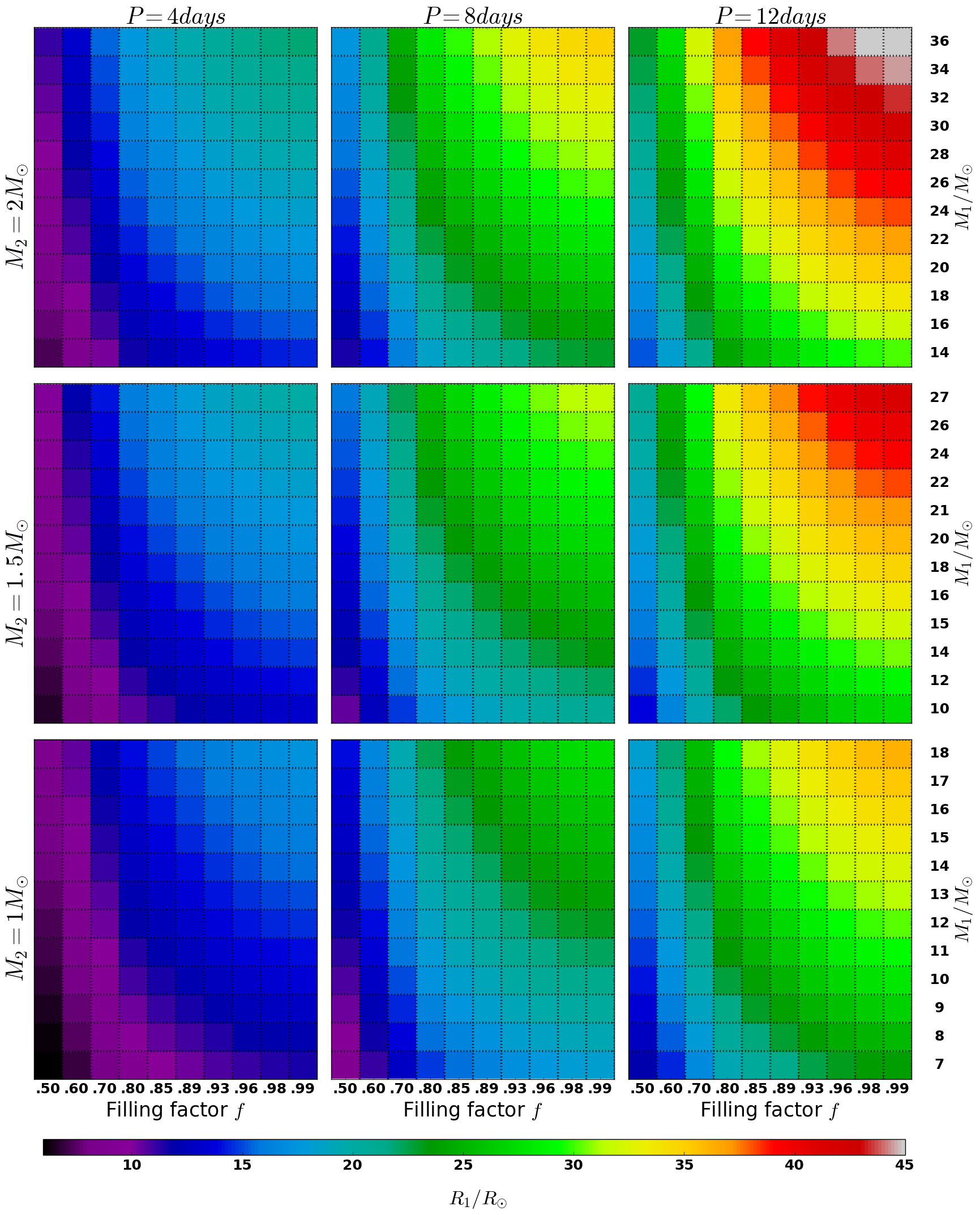}
\caption{All the possible values spanned by the stellar radius $R_1$ as a function of the only parameters which influence it. Indeed, the stellar radius is computed from the filling factor $f$ and the size of the stellar Roche lobe $R_{\text{R,}1}$. Hence the adimensioned $R_1$ depends on the filling factor only but so as to compare it to solar radii, it requires the knowledge of the length scale $R_{\text{R,}1}$, set by the orbital period $P$ (from left to right column), the mass of the compact object $M_2$ (from upper to lower line), the mass ratio (on the right side, already converted into stellar masses using $M_2$) and the resort to Kepler's third law.}
\label{fig:array_R1}
\end{figure*} 

The theoretical four shape parameters our toy-model is based on are theoretical tools which encapsulate all the scale-free dynamics of the problem. However, they do not correspond to the observables we have access to. In this section, we aim at providing a minimal though comprehensive track through the maze of physical quantities in \sgx which fit our model.

We first introduce the relations we will rely on to deduce the broad set of parameters available in this toy-model. We start with the following set of 7 parameters we call prior parameters, selected for their accessibility to observations and/or their requirement to deduce the other parameters :
\begin{enumerate}
\item the orbital period $P$
\item the mass of the compact object $M_2$
\item the stellar temperature $T$
\item the logarithm of the stellar surface gravity $\log (g)$
\item the $\alpha$ force multiplier
\item the $Q$ force multiplier
\item either the terminal speed of the wind $v_{\infty}$ or the average X-ray luminosity of the system $\langle L_X \rangle$
\end{enumerate}
Let us illustrate the procedure we follow starting from the above set of prior parameters. 

Spectroscopic inquiries of the pressure-broadened Balmer and He II lines coupled to stellar atmospheres codes set constrains on the decimal logarithm of the effective surface gravity and on the stellar temperature $T$ \citep[see \eg ][]{Clark2002a} : to know the actual surface gravity, $\log(g)$, thus requires to correct for radiative pressure \citep{Sander2015}. One can however rely on the calibrated data for the same spectral type of stars \citep[see \eg][]{Searle2008,Puls2008}. $\log(g)$ features few dependences ($f$ and $P$, plus a weak dependence on $M_2$ and $q$) such as it narrows the range of filling factors possible :
\begin{equation}
\label{eq:filling}
f \sim \underbrace{\frac{0.32q^{1/2}}{\mathcal{E}(q)\cdot(1+q)^{1/3}}}_\text{78\% $\rightarrow$ 82\%} \left( \frac{M_2}{2M_{\odot}} \right)^{\frac{1}{6}} \left( \frac{9\text{ d}}{P} \right)^{\frac{2}{3}} 10^{\frac{3-\log (g)}{2}} 
\end{equation}
with $g$ in \textsc{cgs} units and where the bracket evaluates the factor it embraces as $q$ goes from 7 to 18. The dependence on the mass of the compact object is also weak, accounting for a few percent only for a neutron star. Since the orbital period is precisely known (when measured), we can affirm that $\log(g)$ is a direct measure of the filling factor. Its precise knowledge (\ie with a precision of at most a few percents), along with the orbital period, is a precious and sufficient asset to enclose the value of the filling factor in a $\sim$10 to 20\% precision range. In parallel, we deduce the Eddington factor from the prior of $T$ and $\log(g)$ :
\begin{equation}
\label{eq:Gamma_roadmap}
\Gamma \sim 0.23 \left( \frac{T}{25\text{kK}} \right)^4 10^{3-\log (g)}
\end{equation}
A first sanity check can be realized here checking whether correcting the effective surface gravity with this value of $\Gamma$ does lead to a value of $\log\left(g\right)$ consistent with the one above, from stellar spectral type considerations. The values we get using the two equations above hold whatever the $\alpha$ and $Q$ force multipliers but will participate to set the mass-loss rate and thus, the average X-ray luminosity.

Concerning the two force multipliers listed as prior parameters, they were calculated by \cite{Shimada1994} for OB-supergiants. For Solar metallicity and effective temperatures between 20 and 30kK, typical of the early-type B stars we consider in the upcoming section, $Q\sim900$ and $\alpha\sim0.5$, with little departure from those values. 

Here, a junction shows up : either we make use of a measured velocity of the wind at infinity\footnote{Which does not depend on the $Q$ force multiplier.}, which is not always available nor precise, or we rely on the X-ray luminosity of the system, which requires a precise estimation of the distance to the \sgx\footnote{Generally available thanks to the spectral energy distribution fits accomplished by \cite{Coleiro2013} and \cite{Coleiro2013a}.} and more dramatically, requires to assume a straightforward relation between the mass accretion rate and the X-ray luminosity (\eg without reprocessing of the X-rays). The first option leads, once the wind speed at infinity corrected for the finite cone effect \eqref{eq:vr_FD} is accounted for, to the following relation :
\begin{equation}
\begin{aligned}
\frac{q}{\mathcal{E}(q)(1+q)^{1/3}} \sim 0.023\frac{f}{1-\Gamma}&\left(\frac{\alpha}{1-\alpha}\right)^2 \times \\ &\left[\frac{v_{\infty}}{(GM_2/P)^{1/3}}\right]^2
\end{aligned}
\end{equation}
The mass ratio is then obtained after a numerical inversion with a Newton-Raphson root finder for instance. This operation however leads to a poor level of precision on the value of $q$ due to a low variation of the \lhs in the range of mass ratios considered. It is thus preferable to make the most of the whole coupling between phenomena our model relies on (not only the coupling between the stellar, orbital and wind properties) and to compare the result obtained above to the one obtained using the X-ray luminosity. The non-trivial coupling of the latter with the other parameters makes a manual monitoring of the $q$ values compatible with the persistent X-ray luminosity observed compulsory but feasible with the web interface aforementioned. The weak dependence of $L_X$ on $q$ blamed in \ref{sec:phys_mass_acc_rate} does not make this operation very precise neither. The matching between the two values of the mass ratio above is a good hint in favour of a reliable value of the mass ratio, what we will show in the next section with eclipsing \sgx (where the mass ratio is observationally constrained).

In any case, once the mass ratio is determined, we are left with all the shape and scale parameters, opening the doors to all the remaining variables and in particular the stellar radius (see Figure\,\ref{fig:array_R1}), mass, luminosity and mass outflow, the wind mass density, the dimensions of the accretion sphere around the neutron star and the circularisation radius compared to the size of the compact object, a clue which indicates the likelihood of the formation of a disk. One can finally summon additional independent measures (\eg of the stellar radius, of the spectroscopically derived stellar mass) or theoretical expectations (\eg the wind momentum - luminosity relationship) to discuss the reliability of the initially considered mass of the compact object, the only truly arbitrary prior.


\subsection{Classical persistent \sgx}

\subsubsection{Selection procedure}

We now discuss the case of three persistent \sgx likely being archetypes of this observational category : Vela X-1, XTE J1855-026 and IGR J18027-2016 (\aka SAX 1802.7-2017). For short, we nickname the second one XTE and the third one SAX. We ended up with these three classical \sgx using the following requirements, suitable to fit our toy-model, applied to the sample of $\sim$20 available \sgx :
\begin{enumerate}
\item an orbital period has been measured.
\item the location of the system in the Corbet diagram \citep[see \eg][Figure 5]{Walter15} is consistent with \sgx (orbital period below 20 days and spin period of the pulsar above 100 seconds).
\item to avoid any discrepancy on the force multipliers due to metallicity-related issues, we focused on \sgx within the Milky Way \citep{Mokiem2007a}.
\item to avoid the net shift in the fraction of the wind being accreted induced by an eccentric orbit - see equation (6) in \cite{Boffin1988} - we selected systems where the eccentricity is below 0.1 \citep[which invalidates for example 4U 1907+097 according to][]{Nespoli2008}.
\item we rejected potential intermediate \textsc{sfxt} such as the eclipsing IGR J16418-4532 and IGR J16479-4514.
\item we rejected the systems where \rlof was suspected or confirmed (Cen X-3).
\item we did not retain the systems hosting a black hole candidate (Cyg X-1 or even 4U 1700-377) due to the mass ratios and the assumption of mildly perturbed spherical wind we considered.
\item we selected systems where the star was not excessively evolved (as in GX301-2).
\item we discarded systems featuring a peculiarly slow wind (OAO 1657-415).
\item the neutron star is eclipsed by its stellar companion so as to be able to retrieve additional information from radial velocity measurements.
\end{enumerate}
Using eclipsing systems where the best-fits ephemerides of the light curves set constrains on the mass ratio is a way to compensate for the low precision of the mass ratio obtained from the measures of the terminal wind speed and the average X-ray luminosity. Nevertheless, the former is flawed by the unknown inclination of the system, an issue our approach is not influenced by : in the incoming Table \ref{tab:paramsAll}, the values between parenthesis are the ones if the system is seen edge-on while the other ones are for the limit-case of \rlof. 

Interestingly enough, a look a posteriori to a Corbet diagram indicates that those three systems lie in a close region of the $P$-$P_{\text{spin}}$ space ; it might support the idea that the properties of the accreting neutron star (age, magnetic field, etc) are similar and indirectly favor a wind dominated mass transfer.

\subsubsection{Observed parameters}

\begin{table*}
\begin{threeparttable}
 \caption{Parameters of the 4 \sgx considered in decreasing order of orbital period. The second set of stellar parameters are deduced from methods not directly based on the stellar spectral type (radial velocity measures, eclipses...). The pairs of numbers for the wind parameters are not the extremal values found in the literature but correspond to the slow and fast regimes we will consider to check self-consistency. The parameters in bold are among the lever quantities we use to deduce the other ones. The X-ray luminosities reported are for an energy range between 3 and 100 keV.}
 \label{tab:paramsAll}
 \begin{tabular}{l|ccc}
  \hline
          & Vela X-1 & XTE J1855-026 & IGR J18027-2016\\
  \hline
   Stellar parameters... & & &\\
   SpT & B0.5Iae & B0Iaep & B1Ib\\ 
   $\mathbf{T/10^3}$\textbf{K} & 25 & 28 & 22\\
   \textit{... from SpT} & & &\\
   $\mathbf{\textbf{log}(g)}$ & 2.9 & 3.0 & 2.7\\
   $R/R_{\odot}$ & 34 & 27 & 35\\
   $\log(L/L_{\odot})$& 5.6 & 5.6 & 5.4\\
   $M_{1\text{,evol}}/M_{\odot}$& 33 & 25 & 22\\
   \textit{... from observations} & & &\\
   $\mathbf{\textbf{log}(g)}$ & 2.9 & 3.1 & 3.1(3.2)\\
   $R/R_{\odot}$ & 32(27) & 22 & 20(17)\\
   $\log(L/L_{\odot})$& 5.6(5.4) & 5.4 & 4.9(4.8)\\
   $M_1/M_{\odot}$& 28(23) & 21 & 19(18)\\   
   Orbital parameters& & &\\
   $\mathbf{P/}$\textbf{day}& 8.96 & 6.07 & 4.47 \\
   $q=M_1/M_2$& 11-13 & 12-16 & 12-16\\
   $M_2$& 1.9(2.3) & 1.4 & 1.3(1.5) \\
   Wind parameters& & &\\
   $v_{\infty}/$\,km$\cdot$\,s$^{-1}$& 700 | 1,700 & 1,600 & 1,100\\
   $\dot{M}/10^{-6}M_{\odot}\cdot$\,yr$^{-1}$& 0.6 | 2 & 1.9 & 1\\
   X-ray luminosity& & &\\
   $D/kpc$& 1.9-2.2 & 9.8-11.8 & 12.4\\
   $F_{14\rightarrow 195}/10^{-11}$\,erg$\cdot$\,s$^{-1}\cdot$\,cm$^{-2}$& 390 & 20 & 7\\
   $F_{17\rightarrow 60}/10^{-11}$\,erg$\cdot$\,s$^{-1}\cdot$\,cm$^{-2}$& 215 & 10.3 & 4.2\\	
   $\langle L_X \rangle/10^{36}$\,erg$\cdot$\,s$^{-1}$& 2.5-3.7 & 3.4-5.1 & 1.9-2.3\\
   & & & \\
   Spin/seconds& 238 & 361 & 140 \\
  \hline
 \end{tabular}
 \begin{tablenotes}
 \small
 \item 1 : \cite{Falanga2015}, 2 : \cite{Coley2015}, 3 : \cite{Clark2014}, 4 : \cite{Quaintrell2003}, 5 : \cite{Coleiro2013}, 6 : \cite{Liu2006}, 7 : \cite{Ducci2009}, 9 : \cite{Prat2008}, 10 : \cite{Hannikainen2007}, 11 : \cite{Walter15}, 12 : \cite{Lutovinov2013}, 13 : \cite{Prinja2010}, 14 : \cite{Searle2008}, 15 : \cite{Mason2011}
 \end{tablenotes}
\end{threeparttable}	
\end{table*}

We searched the literature to gather the parameters displayed in Table \ref{tab:paramsAll}.

The spectral types of Vela X-1 and XTE have been found in \cite{Coleiro2013} who also provide the corresponding effective temperature. For SAX, we used the conclusions of \cite{Torrejon2010}. Following their approach, we read the corresponding stellar parameters in \cite{Searle2008}, Table 5. However, the value obtained from this modeling approach for the surface gravity of SAX leads with equation \eqref{eq:filling} to a star which largely overflows its Roche lobe, for masses of the compact object larger than even 0.5\msun. We thus discard it and pinpoint an important discrepancy between the surface gravity deduced from the stellar spectral type and the one compatible with the accreting regime of the system. Concerning the observed parameters, we deduced the actual surface gravity from the measured radius and mass.

For the observed stellar parameters and mass of the compact object in Vela X-1, we relied on the radial velocity measurements made by \cite{Quaintrell2003}. The values measured by \cite{Falanga2015} essentially agree, and so do the previous values given by \cite{VanKerkwijk1995}. For the terminal speed of the wind, we report both the low value obtained by \cite{Gimenez-Garcia2016} and \cite{VanLoon2001} and the higher terminal speed deduced by \cite{Watanabe2006}. Yet, we notice that \cite{VanLoon2001} and \cite{Gimenez-Garcia2016} consider stellar masses and/or radii marginally compatible with the ones derived by \cite{VanKerkwijk1995}, \cite{Quaintrell2003} and \cite{Falanga2015}. The mass-loss rates are from \cite{Gimenez-Garcia2016} (lower value) and \cite{Watanabe2006} (upper value). The low value of the mass-loss rate and the high value of the terminal speed are consistent with the theoretical predictions for isolated stars of similar spectral type \cite{Searle2008}. 

For the observed stellar parameters and mass of the compact object in XTE, we used the values derived by \cite{Falanga2015}, more precise than the ones derived by \cite{Coley2015}. They based their approach on Monte Carlo best-fit ephemerides of the light curves. However, they tend to determine low inclinations which turn out to be, for XTE and SAX for instance, below the critical ones to avoid \rlof derived by \cite{Coley2015}. Observational biases apart, the filling factor does not physically have to be near unity as previously defended in section \ref{sec:struct_acc_flow}. For the parameters of the wind, since they have not been directly measured, to the best of our knowledge, we used the values listed by \cite{Searle2008} for isolated stars of similar spectral type.

For the observed stellar parameters and mass of the compact object in SAX, we used the values derived by \cite{Coley2015} using radial velocities. As for \cite{Quaintrell2003}, the values between parenthesis are the ones if the system is seen edge-on while the other one is the limit-case of \rlof. The stellar parameters are consistent with those determined by \cite{Mason2011}. \cite{Falanga2015} obtained slightly larger masses for the star and the compact object. For the parameters of the wind, we used the values listed by \cite{Searle2008} for isolated stars of similar spectral type.

We proceeded in the following way to determine estimations of the average X-ray luminosity. We started from the fluxes : we considered both the ones between 14 and 195keV, $F_{14\rightarrow 195}$ observed by the BAT instrument on \href{http://swift.gsfc.nasa.gov/results/bs70mon/}{Swift} (\href{http://swift.gsfc.nasa.gov/results/bs70mon/}{http://swift.gsfc.nasa.gov/results/bs70mon/}), and the ones given by \cite{Walter15} in the hard X-rays, between 17 and 60keV, $F_{17\rightarrow 60}$. We used the estimations of the distances from \cite{Walter15}, \cite{Coleiro2013} and \cite{Kaper1998} for Vela, from \cite{Coleiro2013} for XTE and from \cite{Walter15} for SAX. We neglected the absorption of those sources in these wavebands. To extrapolate to broadband X-ray fluxes, we notice that \cite{Filippova2005} claims a broadband (3 to 100keV) X-ray flux for XTE approximately 3 times larger than $F_{17\rightarrow 60}$ and 1.5 times larger than $F_{14\rightarrow 195}$ ; assuming similar photon energy distribution for the three sources within this energy band, we then deduce the corresponding X-ray luminosities. Those X-ray luminosities are merely estimations which must not be used but as guidelines. The uncertainty we get for XTE and SAX is obtained by using both aforementioned waveband for the flux and by varying the distance to Vela X-1 from 1.8 to 2.2kpc. Notice that for Vela X-1, a shorter distance has also been derived by \cite{Chevalier1998}. Also, the values we get for its average X-ray flux between 3 and 100keV are slightly below the usual X-ray luminosity found in the literature, 4$\cdot10^{36}$\,erg$\cdot$\,s$^{-1}$, possibly because of the higher absorption at the lower limit of the energy range we consider, not corrected here ; those luminosities are thus lower limits. We expect the use of those X-ray luminosities combined with the measures of the terminal speed to yield mass ratios compatible with the ones measured by light curves fitting given in Table \ref{tab:paramsAll}. 

\subsubsection{Self-consistent sets of parameters}

\begin{table}
\begin{threeparttable}
 \caption{Preliminary ranges of filling factors and Eddington parameters (in \%) possible for each of the three systems considered, using the equations \eqref{eq:filling} and \eqref{eq:Gamma_roadmap}. The subscript of each of the filling factor refers to the mass of the accretor, in \msun. It does not affect the value of the Eddington parameter $\Gamma$. The uncertainties represent on one hand the uncertainty on $\log\left(g\right)$, set to 0.05 around the centered value given in Table \ref{tab:paramsAll}, and on the other hand the variation of the mass ratio from 7 to 18. The lower (resp. upper) edge of each range corresponds to $q=7$ (resp. $q=18$) and a maximum (resp. minimum) value of $\log\left(g\right)$. Each time the filling factor obtained was above 100\%, we set a dash to warn the reader that a \rlof star is unlikely.}
 \label{tab:fill_and_gam}
 \begin{tabular}{c|ccc}
  \hline
          & Vela X-1 & XTE J1855-026 &  IGR J18027-2016 \\
  \hline
   $f_1$ & 74 $\rightarrow$ 88 & 80 $\rightarrow$ 96 & 88 $\rightarrow$ -- \\
   $f_{1.5}$ & 79 $\rightarrow$ 94 & 86 $\rightarrow$ -- & 94 $\rightarrow$ -- \\
   $f_2$ & 83 $\rightarrow$ 98 & 90 $\rightarrow$ -- & 99 $\rightarrow$ -- \\
   $\Gamma$ & 22 $\rightarrow$ 38 & 25 $\rightarrow$ 40 & 7 $\rightarrow$ 13 \\
  \hline
 \end{tabular}
\end{threeparttable}	
\end{table}

We start to explore the possible set of self-consistent parameters for Vela X-1 from Table \ref{tab:fill_and_gam}. Given the high value of $\Gamma$, we start by setting it to 30\%. The lower value of the speed at infinity, $700$\kms, requires at most $\alpha=0.45$ and $M_2=1$\msun. The associated X-ray luminosity would be larger than 10$^{37}$erg$\cdot$s$^{-1}$, way above the permanent level observed for Vela X-1. The stellar mean density would also be suspiciously low with a 10\msun star having a 17\rsun radius. Given the observational constrains, we discard this possibility and support the upper measured value of the wind speed at infinity of 1,700\kms. With $\alpha=0.5$ (and $Q=900$), typical of early type B stars with effective temperatures between 20 and 30,000K, we get a more realistic overview. However, even for an accretor mass as high as 2\msun though (where the filling factor has to be larger than 83\%), we hardly reach $v_{\infty}=$1,300\kms, and for very high stellar masses (above 35\msun), hardly compatible with the spectral type. The associated X-ray luminosities would also be excessively large by a factor of 2. Rising $\alpha$ to 0.55 makes the X-ray luminosity drop to a value several times too low and an Eddington factor larger than 30\% is unlikely given the stellar spectral type. On the contrary, keeping $\alpha=0.50$ but lowering $\Gamma$ to 20\%, the low end of the range we deduced in Table \ref{tab:fill_and_gam} from the effective temperature and the surface gravity, brings us to $v_{\infty}\sim$1,600\kms for $q>11$ (and still $f>83\%$). To overlap this area with the available X-ray luminosity and the estimated stellar mass outflow, we must privilege a large value of the filling factor ($\sim 95\%$) and a moderately low value of the mass ratio ($q\gtrsim 12$). Coupled to the observational results derived from the eclipses (\ie $q$ between 11 and 13), it means that Vela X-1 is almost seen edge-on and that the lowest values of stellar masses and radii must be privileged. With a circularization radius one hundred times smaller than the accretion radius, the flow can cicrularize once it reaches a radius of approximately one thousand times the Schwarzschild radius of the accretor. Depending on the extension of the magnetosphere, it might be enough to see a disk-like structure maintained in a short range of radii. The orbital structure of the flow is mostly wind-dominated.

The situation is very similar for XTE which displays a fast wind : a higher velocity means a larger stellar mass\footnote{Remember that the terminal speed scales with the escape speed.} at a similar filling factor and mass ratio, which implies a drop in the X-ray luminosity both because the wind is faster and because the mass ratio is larger. The larger mass outflow is not enough to compensate for the smaller solid angle delimited by the extended accretion sphere. We must thus either consider heavy accretors and set $M_2$ to 2\msun with the canonical $\alpha$ at 0.5 or intermediate mass accretors with $M_2=1.5$\msun but a larger efficiency of the wind launching with $\alpha=0.55$. Notice that the value of $Q$ does not play any role in the value of $v_{\infty}$ and that $\Gamma$ is somewhat larger than the one derived for Vela X-1 (see Table \ref{tab:fill_and_gam}) and can less easily be considered as low as 20\%. If the accretor is heavy, the condition on $v_{\infty}$ enforces a filling factor in a narrow range between 85 and 90\% and a mass ratio above 15 ; if $M_2=1.5$\msun, we are left with all the upper right quarter of the $\left( f,q \right)$ space. In the latter case, the X-ray luminosity is hardly compatible with the observed one, merely reaching 1.5$\cdot 10^{36}$erg$\cdot$s$^{-1}$ for $q=13$ and $f=99\%$, a factor of 2 to 3 below the estimated value. However, for a heavier accretor, the X-ray luminosity found in the region compatible with the measured values of $v_{\infty}$ lies between 4 and 6$\cdot10^{36}$erg$\cdot$s$^{-1}$, in agreement with the observations. We thus support larger stellar masses and radii than the ones displayed in Table \ref{tab:paramsAll} on the basis of this analysis, with a stellar mass loss rate above 4$\cdot$10$^{-6}$\msun$\cdot$yr$^{-1}$. The circularization radius is one hundredth of the accretion radius and one thousand times larger than the Schwarzschild radius of the accretor ; in spite of the large velocity of the flow, it sounds like this configuration still leave room to the possibility to form a disk, alike Vela X-1 which however has a very similar mass ratio and filling factor.

For SAX, Table \ref{tab:fill_and_gam} indicates that we deal with a very compact system where the lower radius of the less evolved star is not enough to compensate for the lower orbital separation : whatever the mass of the compact object, we deal with a stellar companion presenting a filling factor above 90\%. Given the narrow range of $\Gamma$ values centered on 10\%, we enforce it to 10\%. Since we do not expect the star, way more massive than the accretor, to be able to fill its Roche lobe and transfer matter in a stable way, we discard the heaviest accretor at 2\msun and focus on the intermediate and lightest cases. Relying on the measured value of the velocity at infinity at a 10\% precision level, we find that compatible values of mass ratios lie :
\begin{enumerate}
\item below $q=11$, in the lower right corner of $\left( f,q\right)$ space we consider.
\item anywhere between $q=7$ and $q=18$ for $M_2=1$\msun .
\end{enumerate}
Concerning the X-ray luminosity and the stellar mass outflows, the values are systematically lower than the values listed in Table \ref{tab:paramsAll}, mostly due to the lower value of the Eddington parameter of this later spectral type star. Solving this discrepancy requires an anomalously large $Q$ force multiplier, with a larger anomaly for a lower mass accretor. Since the star is less evolved, it is possible that the stellar photosphere is denser which could rise the value of $Q$ by $\sim$20\% \citep[see Table 1 of][]{Gayley1995}, enough to solve the discrepancy for an accretor of 1.5\msun, not for 1\msun. We thus privilege the former option over the latter. For $M_2=1.5$\msun and $Q=1,100$, the configuration with an X-ray luminosity compatible with the observations are confined below $q=12$ and above $f=95$\%, in agreement with the previous argument and in particular with the measure of the wind speed at infinity. The confrontation with the observational constrains inferred using the eclipses shows a marginal overlap with our holistic derivation. If the system is seen close to edge-on, the two approaches match and yield a system with the set of self-consistent fundamental parameters listed in Table \ref{tab:do_it_all} from which all the stellar, orbital, wind and accretion parameters can be derived. In the most likely configuration whose parameters are listed above, the shearing properties are in favor of the formation of a wind-capture disk : the circularization radius is 6 to 10 thousands times larger than the Schwarzschild radius of the compact object and only 10 to 20 times smaller than the accretion radius. It is then likely that a disk forms within the shocked region. The orbital structure of the flow is stream-dominated.

\begin{table}
\begin{threeparttable}
 \caption{Sets of self-consistent fundamental parameters associated to each system. No uncertainty is specified because the systematics dominate and are discussed in the text in more details. Those values serve to compare the three systems to each other, not to specify precisely the parameters of each system. They indicate a trend, not a state.}
 \label{tab:do_it_all}
 \begin{tabular}{c|ccc}
  \hline
          & Vela X-1 & XTE J1855-026 &  IGR J18027-2016\\
  \hline
   $q$ & 12 & 14 & 11 \\
   $f$ & 95\% & 89\%  & 98\% \\
   $\alpha$ & 0.50 & 0.50 & 0.45\\
   $\Gamma$ & $\gtrsim$20\% & 30\% & 10\% \\
   $P$ (days) & 8.96 & 6.07 & 4.47 \\
   $M_2$ (\msun) & 2 & $\lesssim$2 & 1.5 \\
   $Q$ & 900 & 900 & 1,100 \\  
  \hline
 \end{tabular}
\end{threeparttable}	
\end{table}

\section{Discussion and summary}
\label{sec:conclu}
We used a synthetic model to encapsulate the Physics at stake in \sgx and couple the different mechanisms. We showed that the scale-invariant properties can be fully derived from four dimensionless numbers : 
\begin{enumerate}
\item the mass ratio between the supergiant OB star and the neutron star,
\item the stellar filling factor,
\item the $\alpha$ force multiplier associated to the launching of the wind,
\item the ratio of the stellar luminosity to its Eddington luminosity.
\end{enumerate}
In particular, the transition from a stream-dominated flow, reminiscent from \rlof, to a wind-dominated flow is handled continuously and characterized with a threshold on the ratio of the speed of the wind at the orbital separation to the orbital speed. The accurate wind acceleration mechanism we rely on provides a realistic estimation of the kinetic energy the flow acquired when it reaches the vicinity of the accretor. The subsequent accretion amplitudes display a significant enhancement which is enough, in some cases we explicit, to reproduce the observed persistent X-ray luminosities in \sgx in general and in the three systems we considered in particular. If the shearing of the flow is not large enough to produce a disk before the bow shock ahead the \ns \citep{ElMellah2015}, it leaves the door opened to a wind-capture disk within the shocked region, provided the mass ratio is low enough, the wind acceleration inefficient enough (\ie $\alpha$ low) and the filling factor and the Eddington parameter are large enough. In this case, which also favours large X-ray luminosities, the circularization radius is large enough to immune the flow against a premature truncation by the \ns magnetosphere. Finally, the sets of parameters we derive self-consistently for three classical \sgx fit the observed values and, by then, bring additional constrains on the inclination angle, which highlights the robustness of our synthetic model. New astrometric promising instruments are expected to bring much tighter constrains on the orbital parameters of \sgx and the maximum mass of neutron stars \cite{Unwin2008,Tomsick2009,Tomsick2010}, a long-standing quest to understand the equation-of-state of the degenerated matter they are made of.

A by-product that our analysis brought in the spotlight is the suitability of \sgx with a B1-0 type supergiant stellar companion to host a \ns surrounded by a disk-like structure. Indeed, on one hand the mass ratio is low enough to guarantee that a large fraction of the stellar wind is captured and on the other hand, the stellar mass is high enough to reach an effective temperature above the bi-stability jump \citep{Vink1999} and trigger a significant mass outflow ($>10^{-6}$\msun$\cdot$yr$^{-1}$). Since the X-ray luminosity correlates positively with the likelihood to form a disk, it means that as long as the accretor is a \ns, early B donor stars are more likely to see a fraction of their wind forming a wind-capture disk around their compact companion. Even more importantly, the faster and higher mass outflow winds of hotter O supergiant stars (where $\alpha>0.6$) would lead to much lower X-ray luminosities and flow shearing.

Another important feature in our model to reproduce the observed persistent X-ray luminosities is the relatively low values of the $\alpha$ force multiplier required : $\alpha$ between 0.45 and 0.5, while the spectral type of the companion star would suggest $\alpha$ between 0.5 and 0.6. Given the sensibility of the accretion and wind properties to the value of $\alpha$, it is a significant discrepancy. It might be pointing towards an anomalously low metallicity compared to other stars of similar age and spectral type. An alternative to relax this tension without summoning under-metallic composition could be to include additional ingredients in our model susceptible to lower either the wind speed at the orbital separation or to rise the wind density. 

It is suspected that in several \sgx (and \sfxt), wind speeds are sometimes much lower than the terminal speeds given by equation \eqref{eq:vr_FD} in the direction of the accretor \citep{Ducci2010,Romano2015}. A physical process which could bypass the \cak wind acceleration has been proposed by \cite{Hatchett1977}, \cite{Ho1987} and \cite{Stevens1991}. As matter is accreted onto the compact object, the emitted X-rays strip off, upstream in the flow, metal ions of their remaining bounded electrons. As they do so, the wind finds itself no longer able to gain momentum since the fully ionized metal ions can no longer absorb the high energy photons coming from the star. This effect is not covered by our model, mostly because of the underlying positive feedback loop which tends to make systems where this effect prevails switch from a state to another \citep{Karino2014} : as the mass accretion rate rises, the ionization radius rises, slowing down the wind and making the mass accretion rate even higher, until the ionization front goes so far upstream that it inhibates the launching of the wind itself. Since we focus only on persistent \sgx and want to characterize the permanent structure and trends of the flow, we have discarded this effect in the present paper. If we had included it, it would have lowered the wind acceleration efficiency and might have reproduced our results without the need for an excessively low $\alpha$ force multiplier. Its value near the stellar surface would have possibly been in agreement with the models for isolated stars but as the effective opacity of the wind in the resonant metal lines drops because of the photoionizing emission from the flow in the vicinity of the compact object, the value of $\alpha$ would also have dropped.

For the same reason and because we wanted to stick to a ballistic approach to explore the parameters space, we did not consider the inhomogeneities and internal shocks in our model. However, radiatively-driven winds are notoriously unstable and form clumps with overdensities of a factor of 100 or so \citep{Ducci2009}, in particular within a couple of stellar radii from the star \citep{Negueruela2008}, a region where the accretor lies in those close-in \hmxb. Accounting for clumpiness might not just be a question of time-variability and spatial distribution of the wind but might also induce a proper shift of the mean wind density and, by then, of the X-ray luminosity. This point could also help to retrieve the observed persistent X-ray luminosities without having to assume $\alpha<0.5$.

Finally, the prescription we inherited from \cite{ElMellah2015} to extrapolate the fraction of the wind being effectively accreted from the orbital structure of the flow might be an underestimation. A full three dimensional hydrodynamical treatment of the flow within the extended accretion sphere is required to address this question and can now be carried on in a consistent way since the fundamental parameters have been explicited, the associated extension of the vicinity of the compact object have been computed along with the inflow properties (which will serve as outer boundary conditions in mesh-based simulations) and the stream-dominated configurations, susceptible to give birth to a wind-capture disk, have been determined.

\section*{Acknowledgements}
\textsc{iem} wants to thank Allard Jan Van Marle for insightful discussions on the physical mechanism responsible for winds of massive stars and the numerical recipes available to implement it. \textsc{iem} also thanks J\'er\^ome Rodriguez and Alexis Coleiro for their precious suggestions of relevant systems susceptible to be well described by the model presented in this article. Ma\"{i}ca Clavel also brought fruitful comments to evaluate the X-ray absorption of those three systems. 




\begin{small}
\bibliographystyle{agsm}
\bibliography{SgXB_1}
\end{small}




\bsp	
\label{lastpage}
\end{document}